\pdfoutput=1
\documentclass[usenatbib,letterpaper,hyperref]{mn2e}
\usepackage[totalwidth=480pt, totalheight=680pt]{geometry}
\usepackage{amsmath}
\usepackage{color}
\usepackage{natbib}
\usepackage{graphicx}
\usepackage{multirow}
\usepackage{hyperref}

\newcommand{\kmsMpc}{km~s$^{-1}$~Mpc$^{-1}$}
\newcommand{\kms}{km~s$^{-1}$}
\newcommand{\Mpch}{$h^{-1}$~Mpc}

\newcommand{\kpch}{$h^{-1}$~kpc}
\newcommand{\hLsun}{$h^{-2}\;\mathrm{L}_\odot$}
\newcommand{\LCDM}{$\Lambda$CDM}
\newcommand{\healpix}{{\sc Healpix}}
\newcommand{\muKR}{$\mu$K~R$^{-1}$}
\newcommand{\mJy}{$\mu$Jy}
\newcommand{\uK}{$\mu$K}
\newcommand{\mK}{mK}
\DeclareMathOperator*{\sinc}{sinc}
\DeclareMathOperator*{\cotanh}{coth}

\begin{document}

\bibliographystyle{mn2e}

\title[Bulk flows using the CMB]{First measurement of the bulk flow of
  nearby galaxies using the cosmic microwave background}

\author[G. Lavaux, N. Afshordi \& M.~J. Hudson]{Guilhem Lavaux$^{1,2,3,4}$, Niayesh Afshordi$^{1,2}$, Michael~J. Hudson$^{1,2}$\\
$^{1}$ Department of Physics \& Astronomy, University of Waterloo, Waterloo, ON,  N2L 3G1 Canada\\
$^{2}$ Perimeter Institute for Theoretical Physics, 31 Caroline St. N., Waterloo, ON N2L 2Y5, Canada \\
$^{3}$ Department of Physics, University of Illinois at Urbana-Champaign, 1110 West Green Street, Urbana, IL 61801-3080\\
$^{4}$ Department of Physics and Astronomy, The Johns Hopkins University, 3701 San Martin Drive, Baltimore, MD 21218, USA%
}

\maketitle

\newcommand{\hide}[1]{}

\begin{abstract}
Peculiar velocities in the nearby Universe can be measured via the kinetic Sunyaev-Zel'dovich (kSZ) effect. Using a statistical method based on an optimised cross-correlation with nearby galaxies, we extract the kSZ signal generated by plasma halo of galaxies from the Cosmic Microwave Background (CMB) temperature anisotropies observed by the Wilkinson Microwave Anisotropy Probe (WMAP).  Marginalising over the thermal Sunyaev-Zel'dovich contribution from clusters of galaxies, possible unresolved point source contamination, and Galactic foregrounds, we find a kSZ bulk flow signal present at the $\sim 90$\% confidence level in the seven-year WMAP data.  When only galaxies within 50\Mpch{} are included in the kSZ template we find a bulk flow in the CMB frame of  $|V|=533\pm 263$~\kms{}, in the direction $l=324\pm 27$, $b=-7 \pm 17$, consistent with bulk flow measurements on a similar scale using classical distance indicators. We show how this comparison constrains, for the first time,  the (ionised) baryonic budget in the local universe.  On very large ($\sim 500$ \Mpch) scales, we find a 95\% upper limit of 470~\kms, inconsistent with some analyses of bulk flow of clusters from the kSZ. We estimate that the significance of the bulk flow signal may increase to 3-5$\sigma$ using data from the {\sc Planck} probe.
\end{abstract}


\section{Introduction}

Peculiar velocities are the only probe of the large-scale (10 -- 1000 \Mpch\footnote{The parameter $h$ is today's Hubble constant in units of $100$~\kmsMpc. }) distribution of mass in the nearby Universe. Surveys of the peculiar velocity field have recently returned to spotlight, in part due to the availability of larger surveys of galaxy distances, such as the SFI++ \citep{SFIpp} sample of Tully-Fisher distance or the ``First Amendment'' supernova compilation \citep{TH12},
as well as new analysis methods such as the minimum variance weighting method for bulk flows \citep{WFH09} or the Monge-Amp\`ere-Kantorovitch reconstruction \citep{LTMC10}. However, peculiar velocity surveys based on galaxy distances probe scales smaller than $\sim 100$\Mpch{}. The limitation is due to the linear increase of the peculiar velocity error with distance, and the cost of collecting the large samples that are necessary to statistically reduce this error. In contrast, the kinetic Sunyaev-Zel'dovich effect \citep[hereafter kSZ]{SZ80} is sensitive to relative motion of ionised gas with respect to the Cosmic Microwave Background (CMB). Therefore, as the kSZ errors do not depend on distance, it is a promising probe of peculiar velocities on large scales.

Some recent peculiar velocity studies using classical distance indicators have suggested that there may be a cosmic bulk flow on scales of $\sim 50-100$~\Mpch{} in excess of that expected in \LCDM{} \citep{WFH09,LTMC10,FWH10, CMSS11}, whereas other studies have found lower values \citep{NusDav11,TH12}. Clearly, it would be useful to have an independent probe of the peculiar velocity field on these scales.

On much larger scales, the kSZ effect in rich clusters has recently been used to measure the bulk flow by \citet[hereafter KAKE]{KAKE08},  following \citep{KA2000}.  However, subsequent studies using similar data sets \citep{Kei09, OMCP11,MH12} have not confirmed a flow with statistical significance claimed by KAKE. While the results of the latter authors have in turn been questioned \citep{AKEKE10,KAE11,KAE12}, the lack of independent confirmation means that this very-large scale ``Dark Flow'' remains controversial. The main challenge is to disentangle the kSZ signal from the primary fluctuations of the CMB and the instrumental noise. The kSZ signal is at the level of a few $\mu$K, while the CMB is in tens of $\mu$K and the noise may be in hundreds of $\mu$K for WMAP \citep{WMAP_J11}.  Previous kSZ work has focused on clusters of galaxies \citep{HT96,KAKE08,OMCP11,MH12} because clusters hold a significant fraction (but not most) of the ionised plasma in the local universe. Indeed recently \cite{HanAddAub12} have detected the signal of pairwise infall of clusters using the kSZ effect. Unfortunately there are relatively few clusters of galaxies within 100-200\Mpch{} in the nearby Universe, and so, as shown by the studies cited above, the corresponding errors on the bulk flow are large.

In this paper, we use galaxies themselves to produce a kSZ template which we cross-correlate with the CMB fluctuations. Galaxies, including spirals, are expected to have a large plasma halo \citep{FP06, RasSomPed09}, indeed this is a standard assumption of galaxy formation models for over 40 years \citep{ReeOst77, Sil77, WhiRee78}. While much of the hot gas is expected to lie outside haloes, in the so-called warm-hot intergalactic medium \citep[WHIM,][]{DavCenOst01,ShuSmiDan11}, nevertheless, this WHIM gas should be correlated with the haloes of larger galaxies and thus may be detected statistically through cross-correlations \citep[see ][for a claimed detection]{Sol06}.  In the context of this paper, therefore, ``plasma halo'' refers to both the hot plasma in the dark matter halo as well as plasma in the WHIM that is correlated with the galaxy.

The use of galaxies as kSZ tracers has also been advocated by \cite{HDS09} and \cite{SZLJP11}. In particular, using 2MASS galaxies as a probe of the KAKE bulk flow has been proposed by \cite{Z10}.  On the one hand, in a given volume there are far more galaxies than there are clusters of galaxies, e.g.,  within 200\Mpch{}, there are $\sim$60~000 galaxies in the 2M++ galaxy catalogue \citep{LH11}, whereas in the Reflex-eBCS-CIZA catalogue \citep[RBC,][]{KE06}, there are only 273 clusters of galaxies in the same volume. On the other hand, the density profile of the plasma halo is far more uncertain for galaxies than for the clusters of galaxies. Moreover, the plasma halo occupies a much smaller area on the sky, which requires an accurate modelling of beaming effects. Nevertheless there are expected to be more ionised electrons in or near galaxies than in clusters of galaxies, so smaller errors are expected from a galaxy sample.

The basic method used in this paper is to model the observed WMAP maps in all frequencies simultaneously using a combination of foreground templates to represent the kSZ effect as well as the thermal Sunyaev-Zeldovich (tSZ) effect. We also model the contamination from radio point sources that may be present in the galaxies, as well as Galactic foregrounds. We then analytically marginalise all results with respect to primary CMB fluctuations, and additionally marginalise over independent monopoles and dipoles in each channel.

In Section~\ref{sec:stat_method}, we describe the statistical method used to fit the templates. In Section~\ref{sec:templates}, we describe the data and models used to build the templates for the thermal and kinetic Sunyaev-Zel'dovich effects and point source contamination. In Section~\ref{sec:results}, we present and discuss our measurement of the kSZ effect. We compare the kSZ bulk flow with results using other methods and measure the baryon fraction in free electrons in Section~\ref{sec:bulk_flows}. In Section~\ref{sec:discuss}, we discuss future improvements and forecast the errors on the components of the kSZ bulk flow from the upcoming {\sc Planck} temperature maps. Finally,  Section~\ref{sec:conclusion} concludes the paper.


\section{Statistical method}
\label{sec:stat_method}

Here, we adopt a full-sky, multi-channel statistical model for WMAP7 data. We can model the observed sky temperature as a combination of multiple physical effects using a template based method, as in e.g. \citep{GBBHKSW96,JBEGH06}. The templates that we consider in this work will model the two Sunyaev-Zel'dovich effects, the Point Source contaminations, the Galactic foreground emission, the residual temperature monopoles and dipoles. Generically, the expected observed signal $d_c$, in the frequency channel $c$, may be written as:
\begin{equation}
  \mathbfit{d}_c = \mathbf{B}_c \left( \mathbfit{s} + \sum_i \alpha_{i,\sigma_i(c)} \mathbfit{t}_{i,c} \right) + \mathbfit{n},
\end{equation}
where $B_c$ is the beam of the instrument corresponding to channel $c$, $n$ is an instrumental noise component, $s$ is the Cosmic Microwave Background (CMB) signal, $\alpha_{i,\rho}$ are real scalars and $t_{i,c}$ is the template of signal $i$ in the channel $c$ (without beam). We note that $B_c$ is a matrix operation, which can include convolution by any kind of smoothing kernel modelling the instrument response, provided the response is linear.
$\sigma_i(c)$ expresses our prior on the frequency dependence for the physical signal $i$. In the case where the frequency dependence is not known, the operator is the identity, $\sigma_i(c)=c$. This implicitly states that for the template $i$, we have a number of free parameters corresponding to the number of frequency channels. In the case where the frequency dependence of the phenomenon $i$ is known, $\sigma_i(c)$ does not depend on $c$, e.g. $\sigma_i(c)=0$. This states that only one free parameter is required for all channels, e.g. $\alpha_{i,0}$. There may be other intermediate cases if some channels are at the same frequency. We may pick two concrete example that concern us in this work: the tSZ signal and the unresolved point source contamination map. In the case of tSZ, we know the frequency dependence, which is encoded in $t_{\text{tSZ},c}$. We thus only need one parametre $\alpha_\text{tSZ}$. In the case of the map of point sources, the frequency dependence is unclear so we are leaving it free. We will have three parametres: $\alpha_{\text{PSC},Q}$, $\alpha_{\text{PSC},V}$, $\alpha_{\text{PSC},W}$.

We assume that the primary CMB signal is a realisation of a Gaussian random field, and that the instrumental noise is also Gaussian. After marginalisation over the primary CMB fluctuations, we may write the total log-likelihood as (see Appendix~\ref{app:likelihood}):
\begin{multline}
  \chi^2(\{ \alpha_{i,\rho}\}) = \\
  \sum_{c,c'} (\mathbfit{d}_c-\sum_i \alpha_{i,\sigma_i(c)} \mathbf{B}_c \mathbfit{t}_{i,c})^{\dagger} \mathbf{C}^{-1}_{c,c'} (\mathbfit{d}_{c'}-\sum_i \alpha_{i,\sigma_i(c')} \mathbf{B}_{c'} \mathbfit{t}_{i,c'}), \label{eq:chi2}
\end{multline}
with
\begin{equation}
  \mathbf{C}^{-1}_{c,c'} = \tilde{\mathbf{N}}^{-1}_{c} \mathbf{1} \delta_{c,c'} - 
  \tilde{\mathbf{N}}^{-1}_{c} \mathbf{B}_c \mathbf{S}^{1/2} \mathbf{D}^{-1} \mathbf{S}^{1/2} \mathbf{B}_{c'} \tilde{\mathbf{N}}^{-1}_{c'},\label{eq:optimal_weighting}
\end{equation}
and
\begin{equation}
  \mathbf{D} = 1 + \mathbf{S}^{1/2} \left(\sum_{c''} \mathbf{B}_{c''} \tilde{\mathbf{N}}_{c''}^{-1} \mathbf{B}_{c''} \right) \mathbf{S}^{1/2}. \label{eq:D}
\end{equation}
In the above, $c$ and $c'$ run over the available channels of WMAP, $\mathbf{N}_c$ is the instrumental noise covariance matrix in the channel $c$.
The inverse covariance matrix $\mathbf{C}^{-1}$ does not use the noise covariance matrix (only its inverse) and it uses the angular CMB spectrum (and not its inverse). Thus, we do not have to worry about how to regulate the inverse of these two operators. 
In the above, we have used pseudo-inversion and we have set
\begin{equation}
  \tilde{\mathbf{N}}^{-1}_c = \mathbf{M} \mathbf{N}^{-1}_c \mathbf{M} \label{eq:mask_noise},
\end{equation}
with $\mathbf{M}$ the pixel masking operator. In the case of WMAP observations, we assume that the noise covariance matrix is diagonal in pixel space, but we do not assume homogeneity of the noise. It takes the form $\mathbf{N}^{-1}_c=N_{\mathrm{obs},c}/\sigma^2_c$, with $\sigma_c$ the noise normalisation and $N_{\mathrm{obs},c}$ the number of observations for each frequency channels as provided by the WMAP collaboration. The matrix $\mathbf{M}$ is an additional masking operation acting in pixel space. All pixels which are masked are set to zero by this operator. In our case, the mask will correspond to a \healpix{} map, with value either one, for accepting the pixel, or zero, for rejecting it. 
This form has already been used in \cite{WLL04} in the context of Gibbs sampling of the CMB fluctuations.

If we assume that the angular power spectrum of the primary CMB is kept fixed, the equation \eqref{eq:chi2} leads to an analytic solution of the $\alpha_{i,\rho}$:
\begin{equation}
  \alpha_{i,\rho} = \sum_{j,\nu} \mathcal{\mathbf{A}}_{(i,\rho),(j,\nu)}^{-1} \left(\sum_{\substack{c \\ \sigma_j(c)=\nu}} \sum_{c'} \mathbfit{t}_{j,c}^\dagger \mathbf{B}_{c'} \mathbf{C}^{-1}_{c,c'} \mathbfit{d}_{c'}\right) \label{eq:maxlikelihood_alpha}
\end{equation}
with
\begin{equation}
  \mathcal{\mathbf{A}}_{(i,\rho),(j,\nu)} = \sum_{\substack{c,c' \\\sigma_i(c)=\rho, \sigma_j(c')=\nu}} \mathbfit{t}_{i,c}^\dagger \mathbf{B}_{c} \mathbf{C}^{-1}_{c,c'} \mathbf{B}_{c'} \mathbfit{t}_{j,c'}. \label{eq:covariance_A}
\end{equation}

Computing $\mathcal{\mathbf{A}}_{(i,\rho),(j,\nu)}$ requires applying  $\mathcal{O}\left(N_\mathrm{t} \times N_\mathrm{c}\right)$ times the operator $\mathbf{C}^{-1}$. In this paper, we consider ten templates, and for one map per channel, three input maps. This results in applying $\mathbf{C}^{-1}$ thirty times. We do not require more because, for a fixed primary CMB spectrum, we may precompute a weighted template $\tilde{\mathbfit{t}}_{k,c}$, which is sufficient to compute all the terms of the covariance matrix $A$ and the maximum likelihood estimate $\alpha$. The procedure is detailed in Appendix~\ref{app:precomputation}.

For the practical purpose of the numerical implementation, we have separated the fitting of the Galactic foreground templates from the fluctuations due to kSZ, tSZ and point sources contamination. Doing so, the matrix $\mathcal{\mathbf{A}}$, with $21\times 21$ elements, stays the same for all of our experiments on the SZ components and point sources. We have separated the fitting using Gibbs sampling on a set of parameters $\alpha_{i,\rho}$ assuming the other set is fixed. Gibbs sampling provides samples of the posterior distributions. On the positive side, by selecting samples, we can impose a non-Gaussian prior on the  $\{ \alpha_{i,\rho} \}$. On the negative side, we are required to run a sampling chain. If the templates are correlated, this chain may take time to converge. In practice though, for the cases of Section~\ref{sec:results}, the convergence is achieved in a few steps.


\section{Template generation}
\label{sec:templates}

In this section, we describe the physical models used to generate the templates of the three signals considered in this work: kinetic Sunyaev-Zel'dovich (kSZ), the thermal Sunyaev-Zel'dovich (tSZ) and radio point source contamination (PSC). 

The signal from the kinetic Sunyaev-Zel'dovich effect is our primary goal. This is derived from the galaxies belonging to 2M++ galaxy compilation \citep{LH11}, which in turn is based on the 2MASS eXtended Sources Catalog \citep{2MASS} and publicly-available redshift surveys.

The thermal Sunyaev-Zel'dovich (tSZ) signal is produced mostly by clusters of galaxies. As clusters and galaxies are correlated, it is necessary to also model the tSZ signal and marginalise over it. We model the signal using the position and luminosity of clusters in the RBC \citep{KE06} catalogue.

Finally, it may be expected that a subset of local galaxies host radio or submillimetre sources that contaminate WMAP data as unresolved point sources. The population of galaxies and of unresolved point sources may thus be correlated and must also be marginalised over. The contamination by unresolved point sources is obtained using the information contained in the 2MASS eXtended Sources Catalog \citep[hereafter 2MASS-XSC]{2MASS}. 

For fitting the contamination of the temperature fluctuations due to Milky-Way emission, we have used the templates provided by the WMAP7 collaboration on {\sc Lambda}.\footnote{The templates are located at \url{http://lambda.gsfc.nasa.gov/product/map/dr4/templates_get.cfm} .} We have fitted the templates at the full resolution  of $N_\text{side}=512$.

We now describe the templates in detail.


\begin{figure*}
  \includegraphics[width=\hsize]{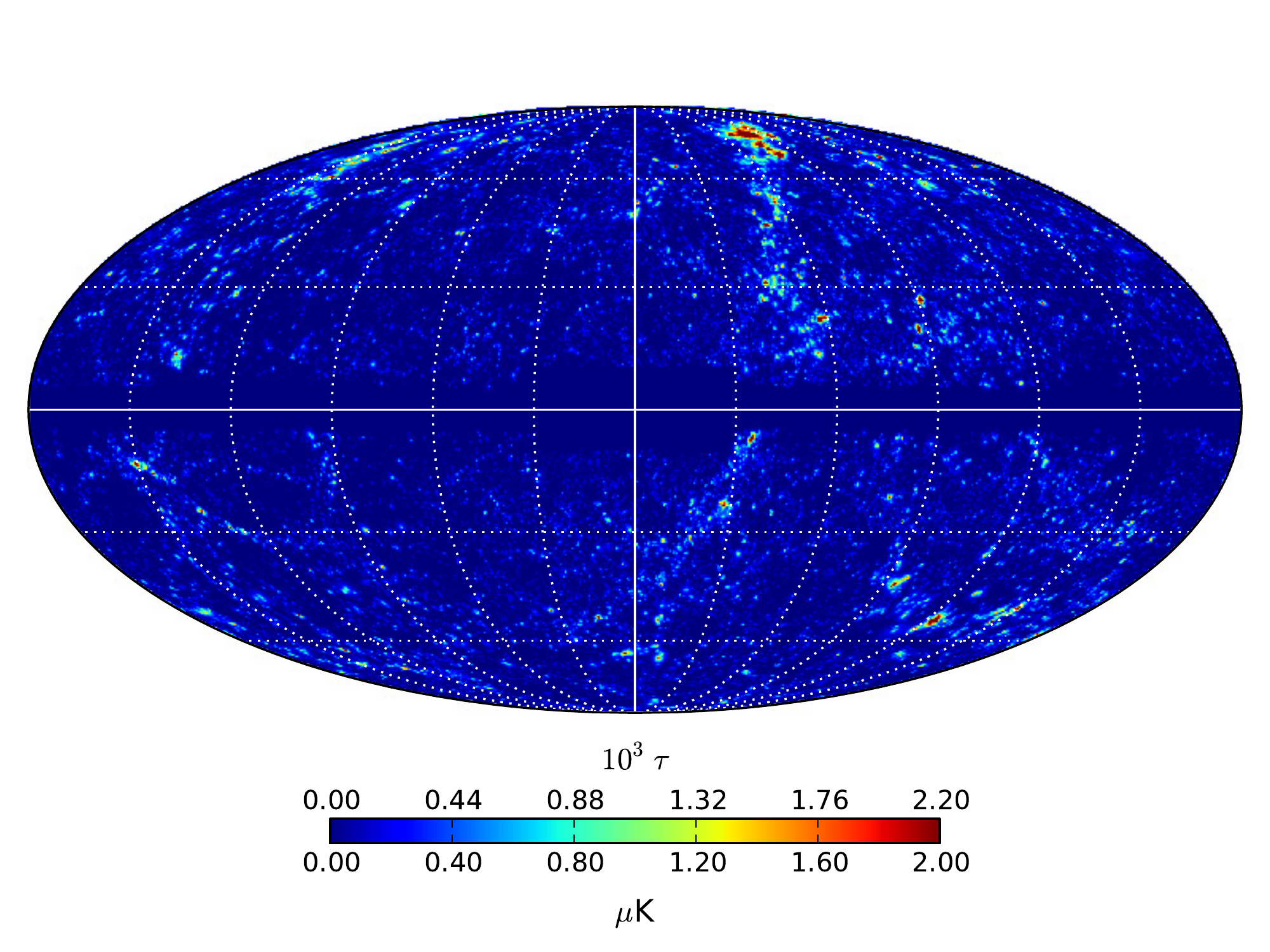}
  \caption{\label{fig:ksz} A map of the expected kSZ signal assuming each galaxy has an inward peculiar velocity of 100~\kms{}, before smoothing with the beam. Only galaxies with $cz \le 20,000$~\kms{} have been included for generating this template. The grid lines are separated by 30$^\circ$ in the longitude and the latitude directions. The color scale also indicates the corresponding optical depth in unit of $10^3$.}
\end{figure*}

\begin{figure*}
  \begin{center}        
    \includegraphics[width=\hsize]{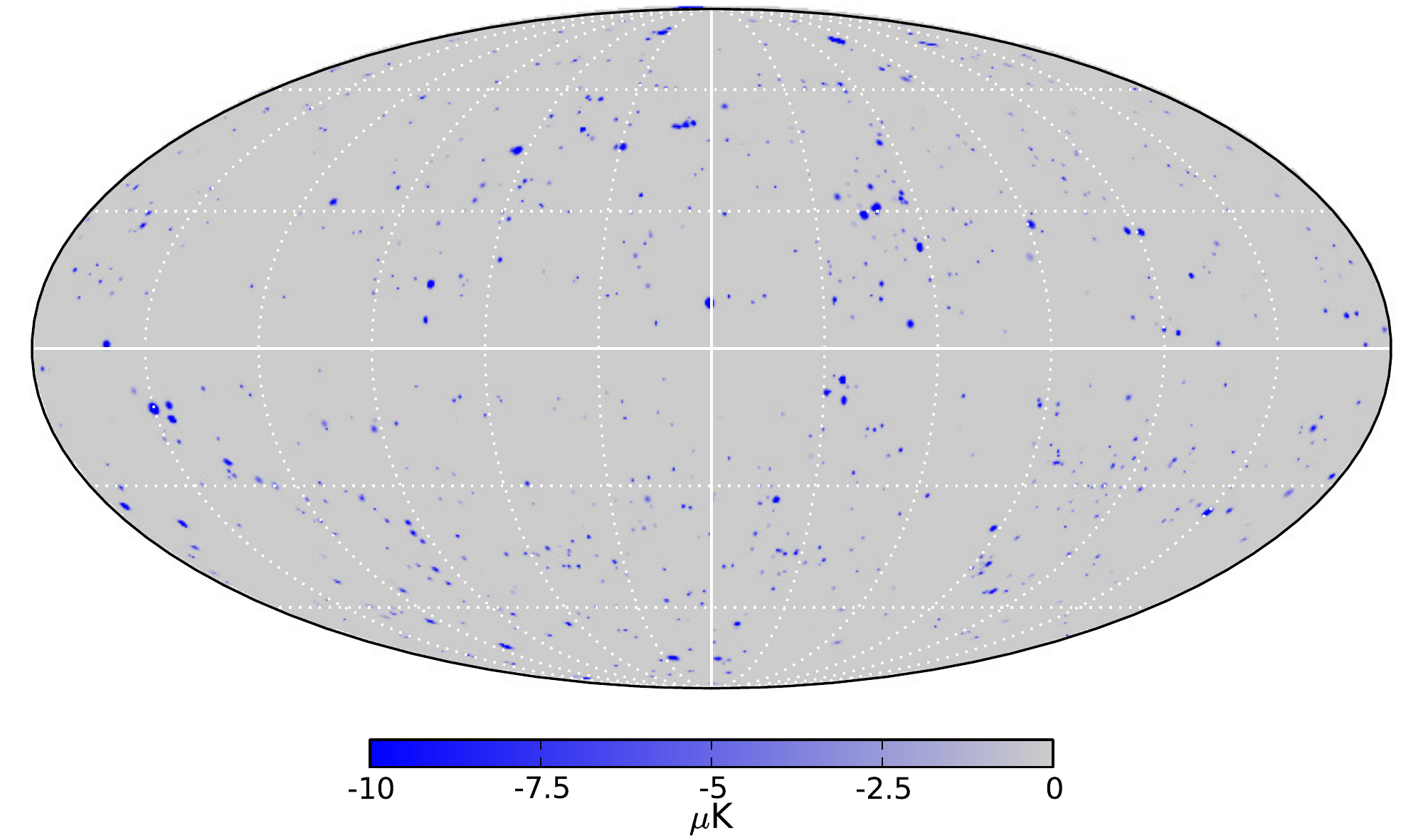}
  \end{center}
  \caption{\label{fig:tsz} A map of the template for the tSZ signal in the W channel of WMAP, before smoothing with the beam. We use a resolution of $N_\mathrm{side}=1024$.}
\end{figure*}

\subsection{kSZ model}
\label{sec:ksz_template}

We assume there is a massive halo of ionised plasma around galaxies. \cite{FP06} proposed an isothermal model for this plasma halo, which we further modify by introducing a hole at the centre of the galaxy. There are at least two reasons to do this. First, we reduce the possibility of correlating the kSZ signal with radio sources located at the centres of galaxies. Second, we expect that the plasma close to the centre of the haloes would cool and collapse to form the cold gas and stars in galaxies. Thus we adopt the following model:
\begin{equation}
  \rho_g(r) = \left\{ \begin{array}{ll} \displaystyle \frac{\Omega_\mathrm{b}}{\Omega_\mathrm{M}} \frac{\sigma^2}{2\pi G} \frac{1}{r^2} & \text{if }\, R_\mathrm{g} < r < R_\mathrm{v}, \\[.5cm]
     \displaystyle 0 & \mathrm{otherwise},
     \end{array} \right.
\end{equation}
with $R_\mathrm{v}$ the maximum extent of our gas profile, and $R_\mathrm{g}=15$\kpch{}. In doing so, we do not attempt to fit the part of the galaxy dominated by stars and may avoid some point source contamination.
The density of electrons is determined from the gas density, and the Helium abundance $Y$
\begin{equation}
  n_e(r) = \frac{1}{m_p}\left(1-\frac{Y}{2}\right)\rho_g(r),
\end{equation}
with $m_p$ the mass of the proton, and $\rho_g$ the total mass density of the gas, and $Y=0.245$ \citep[e.g.][]{WMAP7_Cosmo}.

The kSZ effect depends on the electron density and the velocity of the electrons with respect to the primary CMB, i.e.\ it is proportional to the momentum.  The velocity of a galaxy in the 2M++ sample can be modelled using linear perturbation theory applied to the density field. Here instead, for simplicity, we  model the flow field to lowest order, i.e. by a dipole or ``bulk flow'' with constant Cartesian components $V_i$.
The average kSZ temperature change due to a bulk flow is given by:
\begin{multline}
  \Delta T_\mathrm{kSZ} (\hat{n}) =  - 2\, T_\mathrm{CMB}\, \sigma_T\, \\ \left(\int_0^{\sqrt{(R_\mathrm{v})^2 - r_c^2(\hat{n})}} n_e\left[\sqrt{s^2 + r_c^2(\hat{n})}\right] \mathrm{d}s\right) \left(\sum_{i=1}^3 \frac{V_i}{c} \hat{n}_i\right),
\end{multline}
with $r_c$ is given by Equation~\eqref{eq:rc}.
After integration, we get:
\begin{multline}  
  \Delta T_\mathrm{kSZ} (\hat{n}) =  - \frac{\Omega_b}{\Omega_\text{M}} \frac{T_\mathrm{CMB}\, \sigma_T f_e \sigma^2}{\pi G \mu_e m_p c} \frac{1}{D_A} \\ \tan^{-1}\left(\sqrt{\frac{R^2_\mathrm{v}}{r^2_c(\hat{n})}-1}\right) \left(\sum_{i=1}^3 V_i \hat{n}_i\right) \Theta\left(r_c(\hat{n})-R_\mathrm{g}\right), \label{eq:ksz_model}
\end{multline}
with $\Theta$ the Heaviside step function, $f_e$ being the fraction of free electrons, $\mu_e=2/(2-Y)\simeq1.14$, and $\sigma = 160$~\kms{} the typical velocity dispersion of galactic haloes. $r_c$ is the projected distance from the centre of the galaxy:
\begin{multline}
  r_c(\hat{n}) = D_A(z_c) \sin\left[\cos^{-1}(\hat{n} \cdot \hat{n}_\mathrm{c})\right] = \\ D_A(z_c) \sqrt{1-(\hat{n} \cdot \hat{n}_\mathrm{c})^2 } \label{eq:rc}
\end{multline}
where $D_A(z_c)$ is the angular distance of the galaxy at redshift $z_c$, and $\hat{n}_\mathrm{c}$ is the unit vector in the direction of the galaxy. We set the kSZ template to zero at small radii $r_c(\hat{n}) < R_\mathrm{g}$. Finally, we scale $R_\mathrm{v}$ according to the luminosity as:
\begin{multline}
  R_\mathrm{v} = m_\text{v} \frac{\sigma}{\sqrt{50} H_0} \left(\frac{L_K}{L_*}\right)^{1/3} \\ \simeq  (220 \text{\kpch{}})\, m_\text{v}  \left(\frac{L_K}{L_{*}}\right)^{1/3},
\end{multline}
with $L_{*}=3.8\times10^{10}$\hLsun{} as given for the best Schechter function fit in \cite{LH11}. We take the normalisation from \cite{FP06}. 
The parameter $m_V$ is introduced to increase the effective radius of the plasma halo. As noted above the ``plasma halo'' refers not only to the actual halo of a given galaxy but also to all correlated plasma. The choice of $m_{V}$ is selected so that all free electrons, including those in the WHIM or associated with low luminosity galaxies not in the 2M++ catalogue, are included in the kSZ template (see Appendix~\ref{app:isothermal_sphere}). We adopt $m_V=2.4$ as a default, but consider other possibilities in Section~\ref{sec:systematics}. 

As shown below (Section~\ref{sec:bulk_flows}), for WMAP, we are mostly sensitive to nearby galaxies at a typical distance of 20\Mpch. The typical apparent scale of galaxies on the sky is thus $R_\mathrm{v}/(20$\Mpch$)\sim m_\text{v}\times 0.63$ degrees. For $m_\text{v}=2.4$ ($m_\text{v}=10$ respectively), the scale is thus $\sim 1.5$ degrees ($\sim 6.3$ degrees). In both cases, the modelled plasma halo is bigger than the beam size of WMAP, which is $\la 0.5$~degrees.

We generate three kSZ templates, one for each bulk flow Cartesian axis, and then fit simultaneously all components. In Figure~\ref{fig:ksz}, we show the kSZ template obtained before multiplying by a dipolar bulk-flow and normalised to 100~\kms{}, in the inward direction. Equivalently, it would be the kSZ signal that would be produced if all selected galaxies were moving at the same 100~\kms{}, speed towards us. Using a different colour scale, this same map corresponds to the optical depth. Averaging over pre-beamed non-null pixels, the average optical depth is $1.8\times10^{-4}$ for $m_\text{v}=2.37$. For $m_\text{v}=10$, this same average yields $2.6\times10^{-4}$, a 44\% increase.

The kSZ template exhibits smaller structures than the tSZ signal. Additionally, the kSZ signal does not depend on the frequency. These two features make the kSZ signal far more difficult to detect in the CMB data than the tSZ signal. Nevertheless, we note that the kSZ signal should dominate the tSZ signal on the scale of the plasma halo of a galaxy. The ratio $R_\mathrm{TK}$ between the tSZ signal and the kSZ is given by
\begin{equation}	
	R_\mathrm{TK} = \frac{\Delta T_{\mathrm{tSZ}}}{\Delta T_{\mathrm{kSZ}}} \simeq \frac{2k \bar{T}_e}{m_e V c }
\end{equation}
with $V$ the global velocity of the galaxy, $\bar{T}_e$ the mean temperature of electrons in the halo, $m_e$ the mass of electrons. With $\bar{T}_e = 10^6$~K, $m_e c^2= 511$~keV and $V=300$~\kms{}, we get $R_\mathrm{TK}=0.34$. Additionally, the bulk flow gives a dipolar modulation of the galaxy distribution on the sky, whereas the tSZ signal would simply trace cluster (or galaxy) distribution. Consequently, we do not expect a systematic contamination by tSZ at the galaxy level.


\subsection{tSZ model}

The tSZ signal is generated by hot and dense ionised gas. These conditions are found in the intracluster medium, and clusters of galaxies have already been detected using this effect \citep[for a review see e.g. ][]{B99}.   
We use the model given in the appendix of \cite{ALS05} for the temperature and gas density profiles in clusters. Given the X-ray luminosity, the RBC catalogue provides the virial mass $M_\mathrm{vir}$ from which we derive the virial radius $R_\mathrm{vir} = (2 G M_\mathrm{vir} / (\Delta_\mathrm{vir} H^2))^{1/3}$, with $\Delta_\mathrm{vir}=200$, $G$ the gravitational constant and $H$ the Hubble constant. Using this mass estimate, the model predicts the two profiles assuming hydrostatic equilibrium. The temperature change due to the presence of a cluster in direction $\hat{n}_\mathrm{c}$ is given by:
\begin{multline}
  \Delta T_\mathrm{tSZ} (\nu, \hat{n}) = - 2\, T_\mathrm{CMB}\, \sigma_\mathrm{T} f_{\mathrm{tSZ}}\left(\frac{h \nu}{k T_\mathrm{CMB}}\right) \\ \int_0^{\sqrt{(2 R_\mathrm{vir})^2 - r_c^2}} n_e\left[\sqrt{s^2 + r_c^2(\hat{n})}\right] \frac{k T_e\left[\sqrt{s^2+r_c^2(\hat{n})}\right]}{m_e c^2} \mathrm{d}s,
\end{multline}
with $n_e(s)$  and $T_e(s)$, respectively denoting the density and temperature of the free intracluster electrons at a distance $s$ from the centre of the cluster, $\nu$ the frequency at which the temperature is observed. The dependence $f_\mathrm{tSZ}(x)$ on the frequency $x=h \nu / k T_\mathrm{CMB}$ is
\begin{equation}
        f_\mathrm{tSZ}(x) = 4 - x \cotanh\left(\frac{x}{2}\right).
\end{equation}
$r_c$ is the projected distance from the centre of the cluster. It has the same expression as for galaxies, given in Equation~\eqref{eq:rc}. Note that we stop the integration at twice the virial radius of the cluster.

An example of the tSZ template in the W channel of WMAP is given in Figure~\ref{fig:tsz}. We have selected all 781 clusters with redshift $z \le 0.3$, from RBC. With our normalisation the resulting signal is typically at a level of  a fraction of a milli-Kelvin.


\subsection{Point source model}

We follow \cite{ALS04} and \cite{WMAP_J11} in modelling the effective temperature of point sources in the CMB sky. We remind the reader of their main result:
\begin{equation}
  \Delta T(\hat{p}_j) = \mathcal{N}_{\mathrm{PS}} \Gamma^{\mathrm{bb}}_c a(\nu_c) L_{\mathrm{rad},c} \sum_{i} \frac{\phi_i}{\bar{L}} \delta_S(\hat{n}_i, \hat{p}_j),
\end{equation}
with $\Gamma^{bb}_c$ is a conversion factor between physical flux and antenna temperature that depends on the channel $c$, $\hat{p}_j$ is the direction of the $j$-th pixel, and $\hat{n}_i$ is the direction of observation of the galaxy $i$. We also use $\delta_S(\hat{n},\hat{m})$, which is one if $\hat{n}$ and $\hat{m}$ corresponds to the same pixel of the CMB map, and zero otherwise. $a(\nu_c)$ accounts for the difference between antenna temperature and thermodynamic temperature \citep{WMAP1}:
\begin{equation}
	a(\nu) = \frac{4\sinh^2(x/2)}{x(\nu)^2}.
\end{equation}
$L_{\mathrm{rad},c}$ is the average point source luminosity in the channel $c$.
$\mathcal{N}_\mathrm{PS}$ is a normalisation factor to account both for pixelisation and the normalisation of the beam. $\bar{L}$ is the luminosity density of galaxies in the considered catalogue. We use the typical value derived by \cite{LH11} for normalising: $\bar{L}\equiv 4\times10^{8}$\;\hLsun.  $\phi_i$ is the observed flux of the galaxy $i$:
\begin{equation}
	\phi_i = 10^{0.4 (\mathrm{M}_{K,\odot} - m_i + 25)},
\end{equation}
with $\mathrm{M}_{K,\odot}$ the intrinsic luminosity of the sun in the K band of 2MASS: $\mathrm{M}_{K,\odot}=3.29$. \cite{WMAP1_P03} gives the value for $\Gamma^{\mathrm{bb}}_c$ in the case of free-free emission (scale free spectrum with a slope $\alpha=-0.1$).  The point source template includes only galaxies with apparent magnitude $K_{20} \le 13.5 - A_K$ (with $A_K$ the extinction derived from the reddening \citep{SFD98}). 
The relation between the reddening and the extinction is
\begin{equation}
  A_K = 0.35 E_{(B-V)},
\end{equation}
where the constant of proportionality is obtained from the relation
between extinction in $K$ band and in $V$ band \citep{Cardelli89}.

\subsection{Effective mask}
\label{sec:mask}

Since WMAP data are contaminated by strong radio sources and foreground emission, it is necessary to mask parts of the sky that may contaminate the fitting procedure. We have used the ``Extended Temperature Analysis'' mask (dubbed KQ75 by the WMAP collaboration) to remove Galactic contamination. 

In addition, we  need to allow for the fact that, due to Galactic extinction, galaxies are not uniformly detected on the sky. To ensure the completeness of the 2MASS-XSC sample, and thus of our template, we also require that the Galactic extinction  $A_K < 0.1$. 

\begin{figure}
  \includegraphics[width=\hsize]{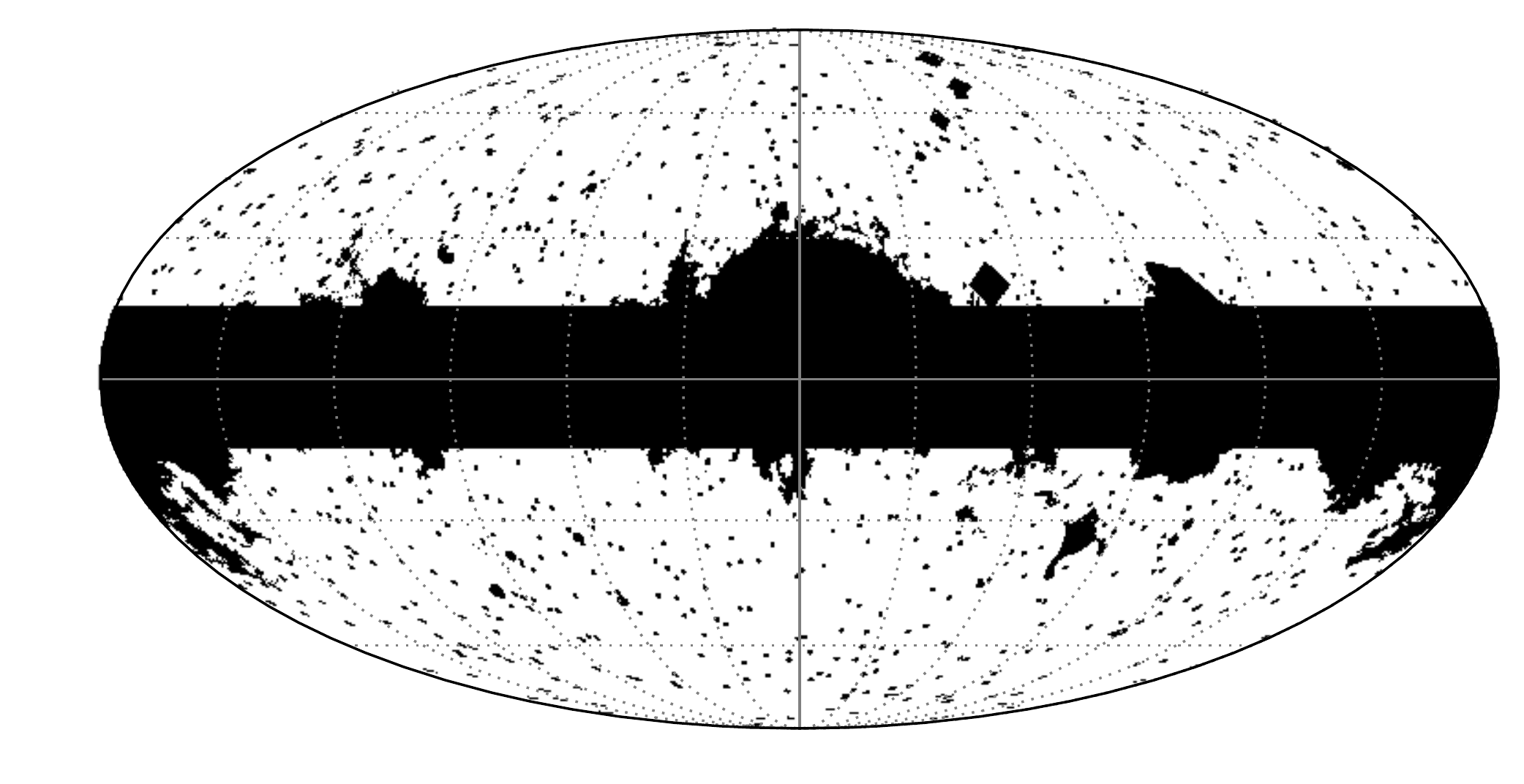}
  \caption{\label{fig:mask} Mollweide projection in Galactic coordinates of one of the masks used in the analysis. This particular  mask is generated from the KQ75 WMAP mask, the constraint that $A_K \le 0.1$ and $|b| \ge 15^\circ$.}
\end{figure}

The 2M++ sample has less uniform sampling 
at low redshift for Galactic latitudes $|b| > 10^\circ$ because at lower latitudes only the galaxies  $K < 11.5$ are properly sampled. We thus add two further conditions: $A_K < 0.1$ and $|b| > 15^\circ$ to produce the effective mask. We use a conservative latitude cut to remove also as much galactic foreground as possible from the fitting. The final mask is shown in Figure~\ref{fig:mask}. The $M$ operator in Equation~\eqref{eq:mask_noise} corresponds to multiplying a pixelated temperature data by this mask.


\begin{figure}
  \includegraphics[width=\hsize]{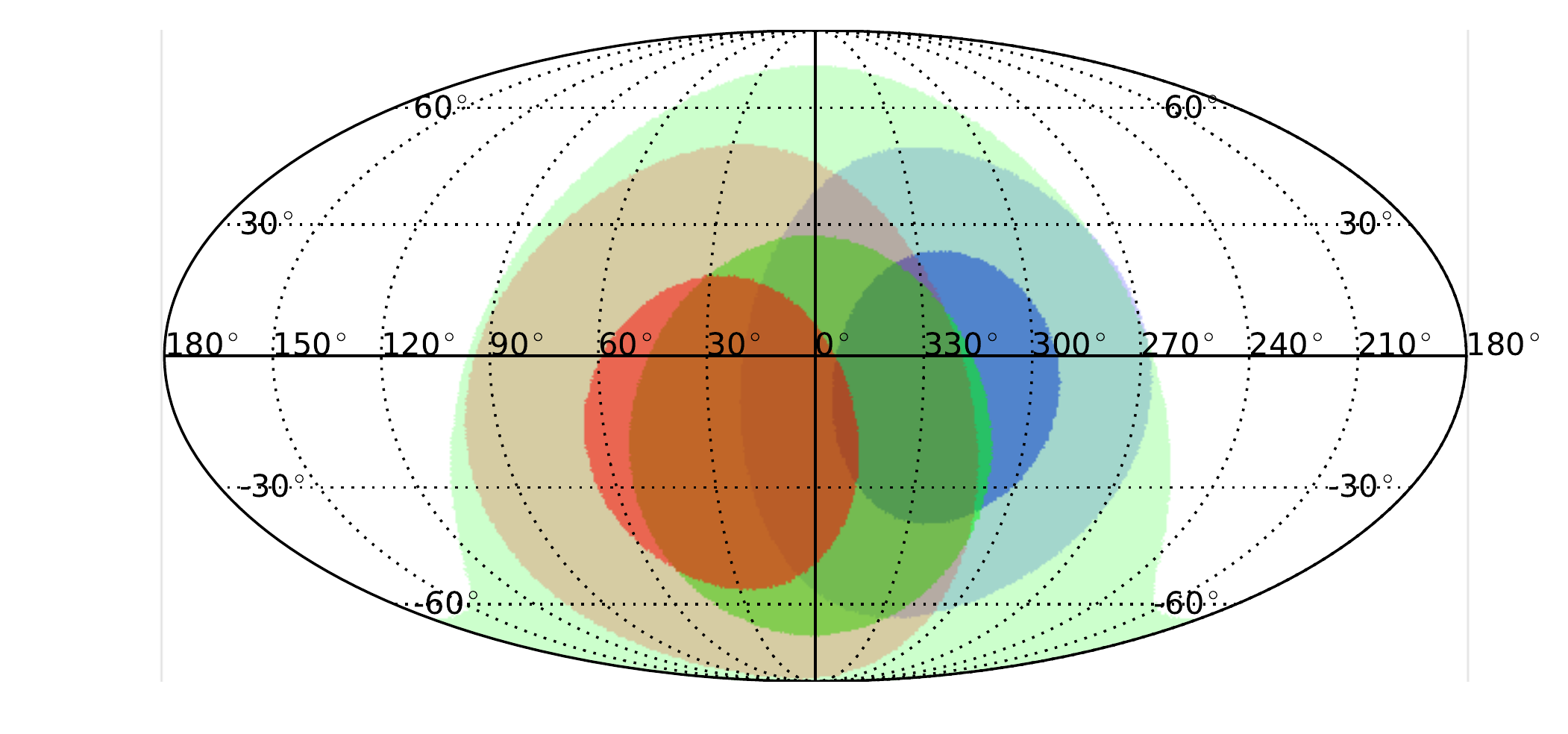}
  \caption{\label{fig:ksz_bulkflow} A Mollweide projection in Galactic coordinates of  the posterior probability of the kSZ bulk-flow direction. The dark (light) areas show the direction of bulk flow at 68\% (95\%) probability. We have only shown the constraints corresponding to the measurement with tSZ signal masked and an isothermal density profile (line 1-3 of Table~\ref{tab:results}).  The blue, green and red areas correspond to a 50\Mpch{}, 100\Mpch{} and 200\Mpch{} respectively, fitting only higher galactic latitudes $|b| \ge 15^\circ$. }
\end{figure}

\section{Results}
\label{sec:results}

In this Section, we present our measurement of the kSZ signal using WMAP7 data. We present the significance and the possible shortcomings of the measurement in Section~\ref{sub:significance}. 


\subsection{Results from 2M++}
\label{sub:significance}

\begin{figure*}
  \includegraphics[width=\hsize]{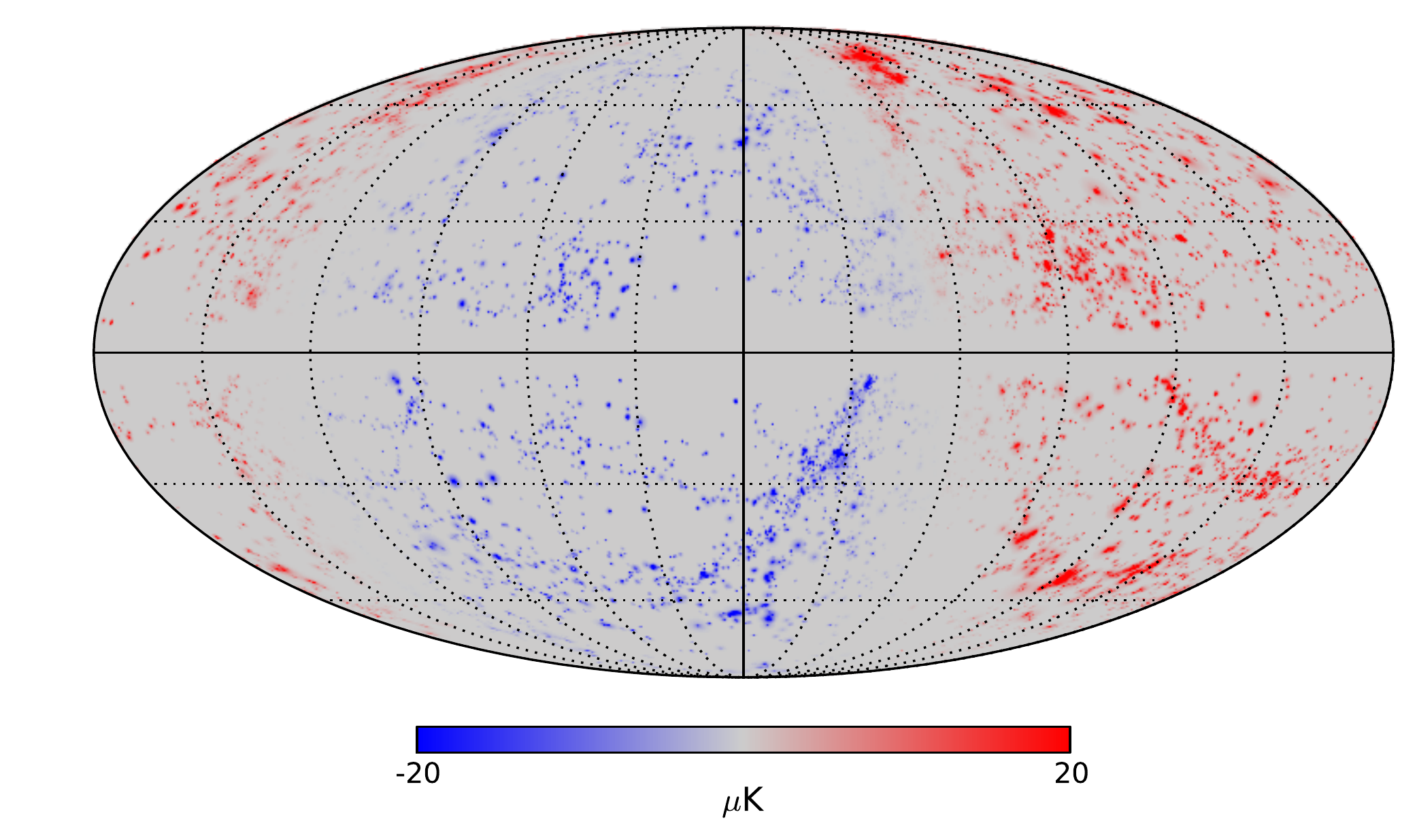}  
  \caption{\label{fig:kszpic} kSZ signal corresponding to the mean bulk-flow corresponding to the 50\Mpch{} cut in the 2M++, limited to $|b|\ge 15^\circ$ (line 1 of Table~\ref{tab:results}). } 
\end{figure*}

Our ``default'' solution is based on the templates and mask described in Section~\ref{sec:templates}, with the amplitude of the templates left free. We use the best fit angular power spectrum provided by the WMAP collaboration \citep{WMAP7_Cosmo} for a \LCDM{} model obtained from WMAP7 data alone \citep{WMAP_J11}. Additionally, we independently fit the residual full sky dipole in WMAP7 maps, which could otherwise contaminate our fit of the kSZ signal, which is also dipolar, albeit localised at the positions of galaxies. We produce a set of kSZ templates with the 2M++ galaxy sample limited at different distances from the observer. All templates discard galaxies within 5\Mpch{} which is the minimum distance at which redshift information can be used as an average indicator of distance. We produce subsamples limited to 50\Mpch{}, 100\Mpch{}, 200\Mpch{} and fit each galaxy sample independently. 

\begin{table*}
  \caption{\label{tab:results} Kinetic Sunyaev-Zel'dovich estimates from WMAP7 maps.}
  \begin{center}
    \begin{tabular}{cccccccc}
      \hline
      Line & Sky cut & kSZ depth & \multicolumn{3}{c}{kSZ} & $\chi^2$ & $p_0$ \\
        & ($|b| \ge$) & (\Mpch{}) & \multicolumn{3}{c}{(\kms{})} & &  (\%)  \\
   \cline{4-6}
         &            &           & $V_x$  &  $V_y$ & $V_z$ \\
      \hline
      1 & $15^\circ$ &  50 &  $559 \pm 290$ & $-412 \pm 285$ & $-93 \pm 200$ & 6.1 & 11\\ 
2 &  &  100 & $305 \pm 223$ & $7 \pm 222$ & $-119 \pm 151$ & 2.5 & 47\\ 
3 &  &  200 & $365 \pm 207$ & $183 \pm 206$ & $-141 \pm 143$ & 5.0 & 17\\ 
      \hline
    \end{tabular}
  \end{center}
  \parbox{\hsize}{{\sc Note:} For all the above measurements, we have used the basic mask of Section~\ref{sec:mask}.
  The reported $\chi^2$ has three degrees of freedom. There is a weak correlation between components that is accounted for in the evaluation of the $\chi^2$. The correlation is 14\% for $V_x \times V_y$, -2\% for $V_x\times V_z$ and 2.6\% for $V_y \times V_z$. $p_0$ is the probability of the null hypothesis, namely that there is no bulk flow. An empty field means that the value is kept the same compared to the previous line. For all measurements we find consistent tSZ measurement with an amplitude of $0.72\pm 0.22$, though the bulk of the tSZ pixels are masked out. Without masking the amplitude of the tSZ is $0.85 \pm 0.08$. The point source contamination signals are, in unit of $10^{-5}$ Jansky, $6 \pm 2$ (Q channel), $12 \pm 7$ (V channel) and $40 \pm 30$ (W channel).
  }
\end{table*}

\begin{table*}
  \caption{\label{tab:robustness} Robustness tests and investigation of model systematics on the kSZ signal detection.}
  \begin{tabular}{clcccccc}
    \hline
    Line & \multicolumn{1}{c}{Experiment} &  \multicolumn{3}{c}{kSZ} & $\chi^2$ & $p_0$ \\
        &            & \multicolumn{3}{c}{(\kms{})} & &  (\%)  \\
    \cline{3-5}
         &            & $V_x$  &  $V_y$ & $V_z$ \\
    \hline
    1 & tSZ masked &  $559 \pm 290$ & $-412 \pm 285$ & $-93 \pm 200$ & 6.1 & 11\\ 
    2 & tSZ not masked &  $510 \pm 294$ & $-239 \pm 269$ & $-183 \pm 204$ & 5.5 & 14\\ 
    3 & NFW, tSZ not masked &  $550 \pm 276$ & $-184 \pm 256$ & $-176 \pm 194$ & 5.4 & 15\\ 
    4 & tSZ not masked, $m_\text{V}=10$       &  $540 \pm 262$ & $-262 \pm 236$ & $-126 \pm 181$ & 6.4 & 10\\ 
    5 & tSZ not masked, $\sigma$ scaling, $m_\text{V}=2.4$     &  $582 \pm 386$ & $-269 \pm 383$ & $178 \pm 302$ & 3.8 & 29\\ 
    6 & tSZ masked, $\sigma$ scaling, limit to $M_*+1$,  $m_\text{V}=3.59$  & $697 \pm 460$ & $-483 \pm 442$ & $506 \pm 345$ & 7.3 & 6\\ 
    7 & tSZ extended masking, $\sigma$ scaling, limit to $M_*+1$,  $m_\text{V}=3.59$ &  $796 \pm 478$ & $-407 \pm 451$ & $429 \pm 352$ & 6.2 & 10\\ 
    8 & tSZ masked, $|b|\ge 10^\circ$ & $531 \pm 273$ & $-405 \pm 259$ & $-122 \pm 194$ & 6.6 & 9 \\
    9 & tSZ not masked, kSZ monopole & $450 \pm 292$ & $-390 \pm 271$ & $-140 \pm 198$ & 5.7 & 13 \\
    \hline
  \end{tabular}\\
  \parbox{\hsize}{
  {\sc Note}: In all the above, we have used a mask based on the KQ75 sky cut, as indicated in Section~\ref{sec:mask}. The notation ``tSZ masked'' indicates that we have masked out pixels which were expected to have un-beamed tSZ signal from the cluster sample. ``tSZ extended masking'' indicates that all pixels with a contamination bigger than 0.1$\mu$K, after beaming in the Q channel, have been masked out. All kSZ templates have been generated from the 2M++ limited to a depth of 50\Mpch{}. }
\end{table*}

Our results are summarised in Figure~\ref{fig:ksz_bulkflow}, Table~\ref{tab:results} and Table~\ref{tab:robustness}.  Our results are consistent the presence of kSZ signal at the 90\% level at line 1. The corresponding kSZ map is given in Figure~\ref{fig:kszpic}. 

 \subsection{Tests of potential systematics}
\label{sec:systematics}

In deriving our solution, we imposed the constraint that that the Galactic foreground corrections are positive. Technically, this corresponds to enforcing the constraint  $\alpha_{i,\rho}>0$, for the templates $i$ corresponding to foregrounds. As a check on the foreground template fitting, we give in Table~\ref{tab:foregrounds} our best fit parameters for the Galactic foregrounds (without forcing the parameter to be positive). We also show in Table~\ref{tab:foregrounds} the value of the parameters found by the WMAP collaboration, which we have averaged by frequency channel. There is a good agreement between the two sets of parameters when we include all pixels at $|b|\ge 10^\circ$. However, there is some tension on the parameters of the $T_K-T_{K_A}$ when only pixels with $|b|\ge 30^\circ$ are included, 
but we note that the WMAP collaboration fitted their models after degrading their map at $N_\text{side}=32$, using {\it all} the pixels of the maps. The cut at $|b|\ge 10^\circ$ is thus more consistent with their methodology. We conclude that our foreground fitting at full resolution is consistent with the findings of the WMAP collaboration.

\begin{table*}
  \caption{\label{tab:foregrounds}%
    Foreground contamination of the map estimated from the data.}
  \begin{center}
    \begin{tabular}{cccccccccc}
      \hline
      Origin  & \multicolumn{3}{c}{Dust map}  & \multicolumn{3}{c}{$T_\mathrm{K}-T_\mathrm{KA}$ map} & \multicolumn{3}{c}{$H_\alpha$ map} \\
              & \multicolumn{3}{c}{ }         & \multicolumn{3}{c}{ }  & \multicolumn{3}{c}{($\mu$K~R$^{-1}$)} \\
      \cline{2-10}
              & Q & V & W &  Q & V & W                   & Q & V & W \\
      \hline
$|b| \ge 10^\circ$ & $0.4 \pm 0.2$ &$0.6 \pm 0.2$ &$1.3 \pm 0.2$ &$0.32 \pm 0.02$ &$0.12 \pm 0.02$ &$0.06 \pm 0.02$ &$1.7 \pm 0.4$ &$1.2 \pm 0.4$ &$0.8 \pm 0.4$  \\
$|b| \ge 30^\circ$ & $-0.6 \pm 0.3$ &$-0.4 \pm 0.3$ &$0.3 \pm 0.3$ &$0.42 \pm 0.02$ &$0.23 \pm 0.02$ &$0.17 \pm 0.02$ &$1.8 \pm 0.4$ &$1.1 \pm 0.4$ &$0.7 \pm 0.4$  \\

      WMAP7 &   0.20 & 0.47 & 1.28  & 0.23 & 0.047 & 0.00 & 1.22 & 0.78 & 0.43 \\
      \hline
    \end{tabular}
  \end{center}
    \parbox{\hsize}{  {\sc Note}: We show the best fit amplitude for each of the templates provided by the WMAP collaboration \citep{WMAP_Foreground_11}, at a resolution of $N_\text{side}=512$. The first two lines has been obtained in this work, using the indicated sky cuts. The last line has been derived from table~2 of \protect\cite{WMAP_Foreground_11}.}
\end{table*}

After subtracting the Galactic foreground components, we are left with a signal for point-sources which is consistent with zero for the V and W channels and marginally above zero for the Q channel. The tSZ template has an amplitude of $0.85\pm 0.08$, significant at $\sim 10.6\sigma$, and consistent with unity. A value of unity means that the physical model adopted for the tSZ effect is exactly right.

In Table~\ref{tab:robustness}, we investigate how different cuts or models affect the results of line 1 of Table~\ref{tab:results}. The first line of this table is a copy of the first line of Table~\ref{tab:results}. We check the impact of pixel masking in line 2. For this experiment, we use all the pixels of the map and we assume that our model for the tSZ signal is correct in average. This operation introduces a marginal change in the kSZ signal. 

Our default model for the plasma halo discussed in Section \ref{sec:ksz_template} is simple and it is possible that the plasma traces halos of different mass differently, or has a different radial dependence.  In lines 3 to 6, we investigate the impact of the choice of the model  for the density profile of the plasma halo. Using a Navarro-Frenk-White profile \citep{NFW96} instead of an isothermal profile barely changes the result (line 3 vs.\ line 1). We have attempted to scale up the plasma halo radius $R_\text{V}$ by changing the multiplicative factor $m_\text{V}$ to the extreme value $m_\text{V}=10$. Increasing this factor may be useful for including the presence of smaller haloes and/or filamentary WHIM that are clustered with the galaxy. This operation does not affect mean recovered velocity and slightly reduces the error bars by introducing more pixels in the correlation. This suggests for WMAP that it is difficult to measure the radial extent of clustered electrons via the strength of the kSZ effect 

It is also possible that the mapping between stellar mass traced by $K$-band light and plasma halo mass  differs from our fiducial choice for which the mass is proportional to $L^{1/3}$. In lines 5 to 7, we have introduced the scaling of the velocity dispersions with the luminosity of the galaxy according to
\begin{equation}
  \sigma(L) = 160~\text{km s}^{-1} \left(\frac{L}{L_*}\right)^{1/3}\,,
\end{equation}
where  $L_{*}=3.9\;10^{10}$\hLsun. This choice effectively forces mass to be proportional to the luminosity. The first impact of such a choice is just to increase error bars (line 5), which is expected as the kSZ signal from small galaxies is strongly reduced. Limiting the sample to the 4450 galaxies within 50\Mpch{}  which have an absolute magnitude brighter than $M_*+1$ and adjusting $m_\text{V}$ accordingly (Appendix \ref{app:isothermal_sphere}) leads to a strong change in the flow amplitude along the $Z-$direction (line 6).  The cause of this change is not understood. In line 7, we have tried to remove as much tSZ signal as possible by masking out pixels which were contaminated at a level larger than 0.1~$\mu$K after beaming according to our model and the RBC cluster catalogue. The kSZ signal does not change compared to line 6.

In line 8, we have reduced the masked regions to include more temperature measurements at lower galactic latitudes. Compared to the fiducial experiment in line 1, the error bars are slightly smaller in all directions. The mean values per directions are in perfect agreement with the values in line 1.

For the case 9, we consider whether we may miss a kSZ monopole component and/or a tSZ component generated by the plasma halo around galaxies.  Not accounting for this effect in our model may bias the results. To assess this effect, we add a new template that corresponds to a kSZ monopole. We find a monopole for the kSZ template equal to a mean (inward) peculiar velocity of  $-243 \pm 152$~\kms{}. The $\chi^2$ value takes into account errors correlation between the component, which are introduced by the mask.

\begin{figure}
  \includegraphics[width=\hsize]{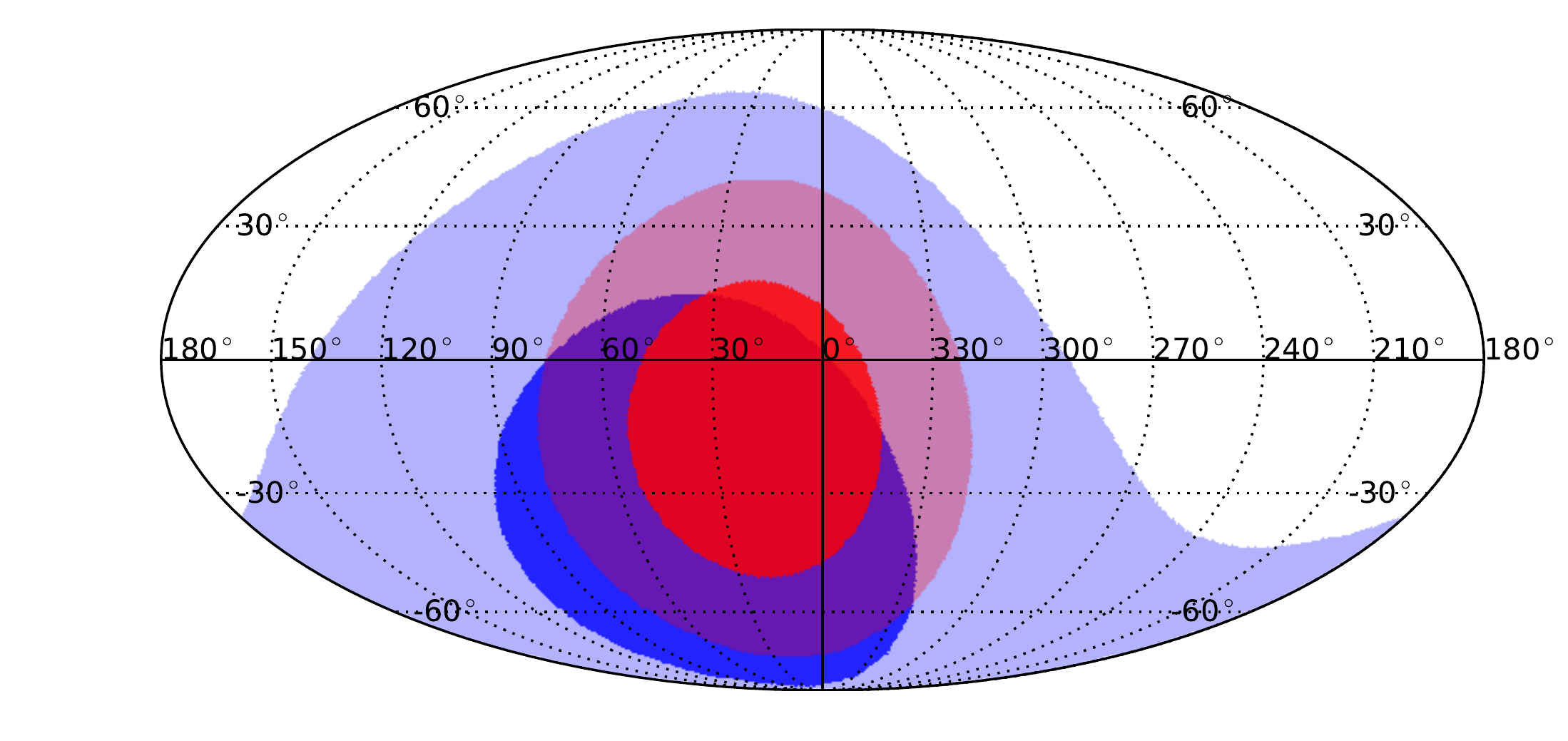}
  \caption{\label{fig:galactic_plane_influence} Test of contamination of the kSZ signal by the Galactic plane. We show the direction of the fitted kSZ bulk flow in two cases where we only use pixels with  $|b|\ge 10^\circ$ ($|b|\ge 30^\circ$ respectively) for the red (blue respectively) contours. In both cases, we have limited our galaxy sample for generating the kSZ templates to 200\Mpch{}.}
\end{figure}

In Figure~\ref{fig:galactic_plane_influence}, we show the likelihood contours of the direction of the kSZ signal for two different mask cuts $|b|\ge 10^\circ$ and $|b|\ge 30^\circ$. The test is run for a kSZ depth of 200\Mpch. There is no sign of strong variation of the peak of the likelihood by cutting $\sim$40\% of the available pixels. The only visible impact is a widening the distribution.

\begin{figure*}
  \includegraphics[width=\hsize]{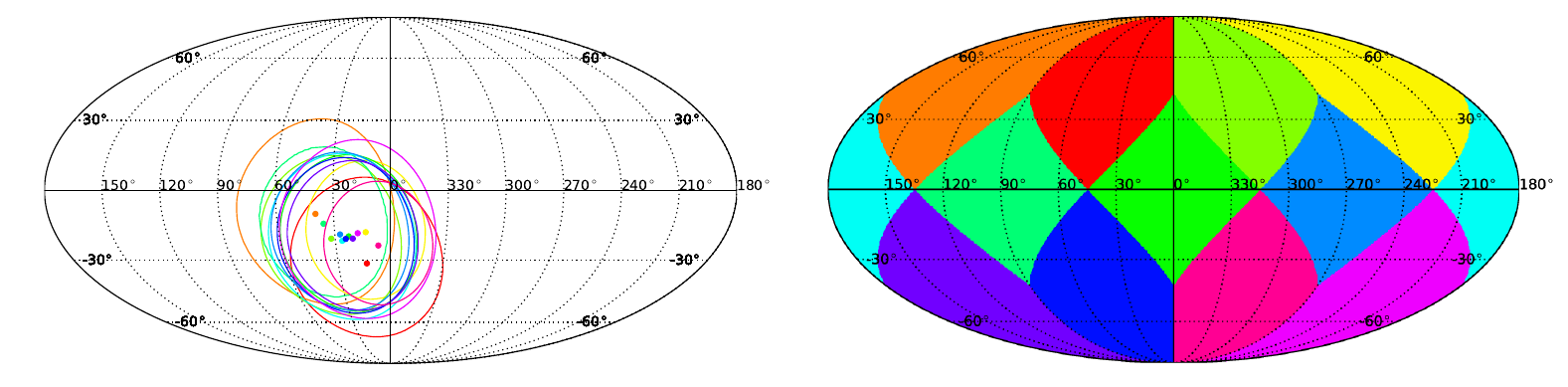}
  \caption{\label{fig:jackknife} In the right panel, we show the change in direction of the flow resulting from a jackknife test. We removed the indicated root pixel of the \healpix{} mesh for each test. We have used a low-Galactic latitude cut of the mask with $|b|\ge 10^\circ$. The kSZ template has been generated with galaxies within a 5-200\Mpch{} from the Local Group. The line iso-density contour gives the area corresponding to 68\% of probability. The root pixels of the \healpix{} mesh are given in the right panel. }
\end{figure*}

Finally, in the left panel of Figure~\ref{fig:jackknife} we show the result of a jackknife test obtained by removing the part of the sky that corresponds to the indicated root pixel of the \healpix{} mesh. These root pixels are given in the right panel of Figure~\ref{fig:jackknife}. This test shows that, at a given depth, the kSZ signal is stable. 
We note that for deeper measurements, the jackknife causes the recovered flow to scatter along the X-direction and thus could be interpreted as a sign of contamination by the WMAP ``haze'' \citep{F04,PGB11}. Correlations between galaxies and Milky-Way extinction properties are likely in the bulge and that could correlate the kSZ signal with the haze. We note that the effect of this haze would go in the opposite direction as what we obtain for the kSZ in all slices: it creates a positive temperature fluctuation which would push the kSZ signal towards negative X. In our case, the signal is pushed towards positive X. Additionally, the push would be affected by removal of the the galactic bulge part in the Jackknife test, which is not the case. 

\subsection{Results from 2MASS-XSC}

\begin{table*}
  \caption{\label{tab:deep_flow} Bulk flows at depths $> 200$ \Mpch{} from the kinetic Sunyaev-Zel'dovich effect alone.}
  \begin{center}
    \begin{tabular}{ccccccccc}
      \hline
      Experiment & Typical depth & \multicolumn{3}{c}{kSZ} & $\chi^2_0$ & $p_0$ & $\chi^2_\text{KAEEK}$ & $p_\text{KAEEK}$ \\
                 & (\Mpch)       & \multicolumn{3}{c}{(\kms{})} &     & (\%) & & (\%) \\
      \cline{3-5} 
                 &               & $V_x$ & $V_y$ & $V_z$ & & \\
      \hline
      2MASS-XSC $K\le 14$ & 470 &  $160 \pm 141$ & $211 \pm 124$ & $-11 \pm 112$ & 4.3 & 23 & & \\
      \cite{KAEEK10} & 175-260   &  $174 \pm 407$ & $-849 \pm 351$ & $348 \pm 342$ & 7.1 & 7 & 9.1 & $2.8$\\
                     & 260-380   &  $428 \pm 375$ & $-1029 \pm 323$ & $575\pm 316$ & 14.7 & 0.2 & 16.3 & $9\,10^{-2}$  \\
                     & 270-530   &  $352 \pm 304$ & $-713 \pm 262$ & $652 \pm 256$ & 15.2 & 0.16 & 16.1 & $1\,10^{-1}$ \\
      \hline
    \end{tabular}    
  \end{center}  
  \parbox{\hsize}{{\sc Note:} $\chi^2_0$, and its associated probability $p_0$, represent the probability that the flow is zero. $\chi^2_\text{KAEEK}$, and its associated probability $p_\text{KAEEK}$, is the probability that the results of \cite{KAEEK10} are in agreement with our result based on 2MASS-XSC. }
\end{table*}

Finally, we have estimated the amplitude of the bulk flow using the entire 2MASS-XSC limited to  $K_\mathrm{fe} \le 14.0$ as the kSZ template.  After removing nearby large galaxies and parts of the sky that could suffer strong reddening, we have 959~554 galaxies with $K_\mathrm{fe} \le 14.0$. As the 2MASS-XSC does not have redshifts, we randomly generated distances for each galaxy based on their apparent magnitude and an assumed Schechter function for the distribution of absolute magnitudes of galaxies. Specifically, we used the Schechter parameters given in \cite{LH11} for the 2M++: $\alpha=-0.73$, $M_*=-23.17$. The result is not strongly dependent on these parameters. We estimated the kSZ bulk flow from an ensemble of ten realisations of those distances. The scatter in bulk flow from one realisation to the other is  $\sim 60$~\kms{} by component. The median depth of such a survey is 470\Mpch{}, with a mean distance of $\sim$550\Mpch{} because of the long tail of the distribution. The results are given in Table~\ref{tab:deep_flow} and are consistent with no bulk flow on these very large scales, with a 95\% upper limit of 520~\kms, after correction for random error biasing. The standard deviation per component of the kSZ signal is $\sim$100~\kms{}.  We have not found any evidence for a bulk flow that would be inconsistent with \LCDM{}.  However, our results disagree with the bulk flow found by \cite{KAEEK10} at the 4.2$\sigma$ level \footnote{A similar conclusion was reached in \cite{2011JCAP...04..015D} through analysis of supernovae Ia at $z<0.05$.} \footnote{After initial submission of this paper, \cite{LZC12} published an attempt to detect the ``Dark Flow'' using a cross-correlation of SDSS galaxy positions and WMAP temperature measurements. While the authors found some evidence of a correlation, they acknowledge that they were not able to reliably link it to a large-scale bulk flow.}.


\section{Bulk flows and the fraction of free electrons}
\label{sec:bulk_flows}

While the bulk flow on very large scales is controversial, the existence of a bulk flow at depths of $<50$ \Mpch{}, due to local superclusters such as Virgo, the Great Attractor region and Perseus-Pisces, is not controversial: there is good agreement between peculiar velocity surveys as well as with the predictions of the density field.  The kSZ signal is the product of the electron density times the velocity, i.e. the momentum. Therefore, we can take the existence of the bulk flow as a given, and instead determine the density of free electrons. 

In order to do so, we must first consider the volume that is actually being sampled by combination of the kSZ template and the WMAP data. Although our template extends to 200 \Mpch{},  nearby galaxy haloes have more weight, and so the bulk flow measured is not simply volume limited. We discuss the effective volume of the kSZ measurement below before discussing the fraction of free electrons.

\subsection{Kinetic Sunyaev-Zel'dovich effect window function}

\begin{figure}
  \includegraphics[width=\hsize]{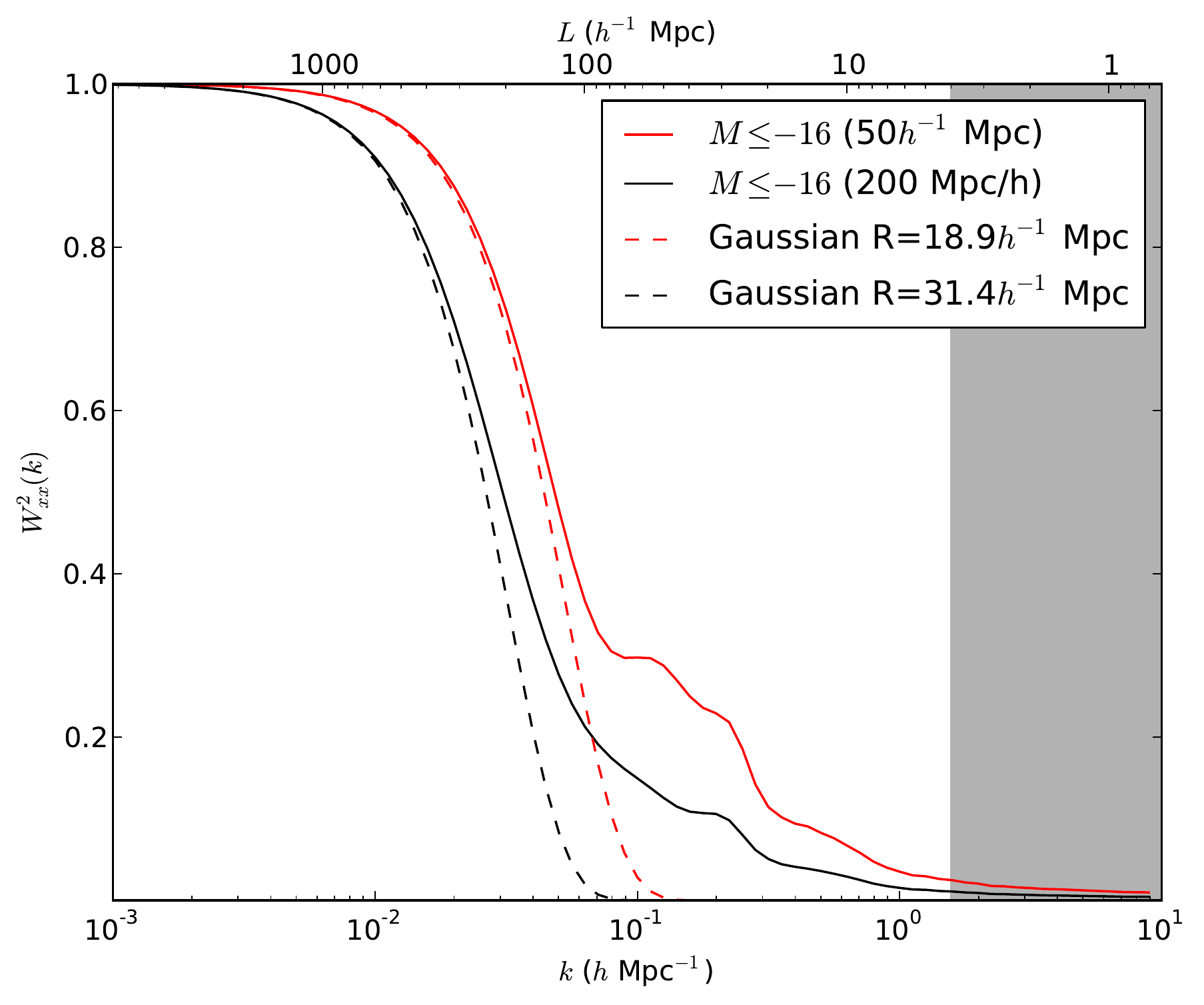}
  \caption{\label{fig:window_function} We show the window function $\mathcal{W}^{x,x}(k)$ (component along the $X$ axis of the bulk flow)  for our two sample selections: 50\Mpch{} (red lines) and 200\Mpch{} (black lines). They have been derived using Eq.~\eqref{eq:window_prefinal} of Appendix~\ref{app:ksz_window}. We show in dashed lines the fit to a Gaussian window function to the corresponding window functions. The part of the window function highlighted by the greyed area may not be trusted because of the approximations that we have used.}
\end{figure}

The amplitude and direction of the kSZ effect may be interpreted as a bulk flow obtained from the weighted average of the peculiar velocities of galaxies. The weighting depends both on the selection cut and the limitations of the instrument used to observe the CMB. The weighted average procedure may be modelled by the convolution of the peculiar velocity field by a kernel, called window function. We argue that the bulk flow estimate must have a similar window function to that used in a maximum likelihood analysis of galaxy distance surveys. We may assume, e.g., that galaxies have all the same physical size $\sim R_\mathrm{v}$. Their sizes on the sky, in terms of number of pixels, are thus $\sim R^2_\mathrm{v}/z^2 \propto 1/d^2$, with $z$ their redshifts and $d$ their angular distances. The number of pixels affected by each galaxy in the kSZ template is thus scaling as $1/d^2$. If for simplicity we assume that the pixel noise is homogeneous, the weight for the peculiar velocity of each galaxy becomes proportional to the inverse squared distance of this galaxy. This is exactly the same case for bulk flow derived from maximum likelihood estimate from galaxy distances. In that case, the peculiar velocity errors scale as the distance $d$, which leads to the same weight per galaxy as the kSZ effect.  

We give a quantitative derivation of this window function in Appendix~\ref{app:ksz_window} where we also discuss the limitations of this calculation. 
Figure~\ref{fig:window_function} shows the square of the window function of the bulk flow in the Galactic bulge direction ($X$ axis).
We also show the window function corresponding to the convolution with a Gaussian kernel $W_\mathrm{G}$ of size $R$ defined as 
\begin{equation}
        W_\mathrm{G}(\mathbf{r}) = \frac{1}{(2\pi)^{3/2} R^3} \text{e}^{-\frac{r^2}{2 R^2}}.
\end{equation}
The size $R$ is determined such that the second derivatives of the actual window function and of the Gaussian kernel are equal. The measurement, based on galaxies within 50\Mpch{}, seems to correspond to a $18.9$\Mpch{} Gaussian window on large scales. Similarly, the measurement using galaxies within 200\Mpch{} corresponds to a Gaussian window of 31.4\Mpch{}. We have evaluated the window function for the Galactic $Y$ and $Z$ components of the bulk flow and we have obtained similar results for the probed scales. Because the masked galaxy distribution is anisotropic, the window functions are not strictly equal for the three components of the bulk flow.  The important point is that the kSZ signal is dominated by galaxies on scales significantly nearer than the limiting depth.

\subsection{Comparison with published bulk flows and the free electron fraction}
\label{sec:comparison_bf}

\begin{figure}
  \includegraphics[width=\hsize]{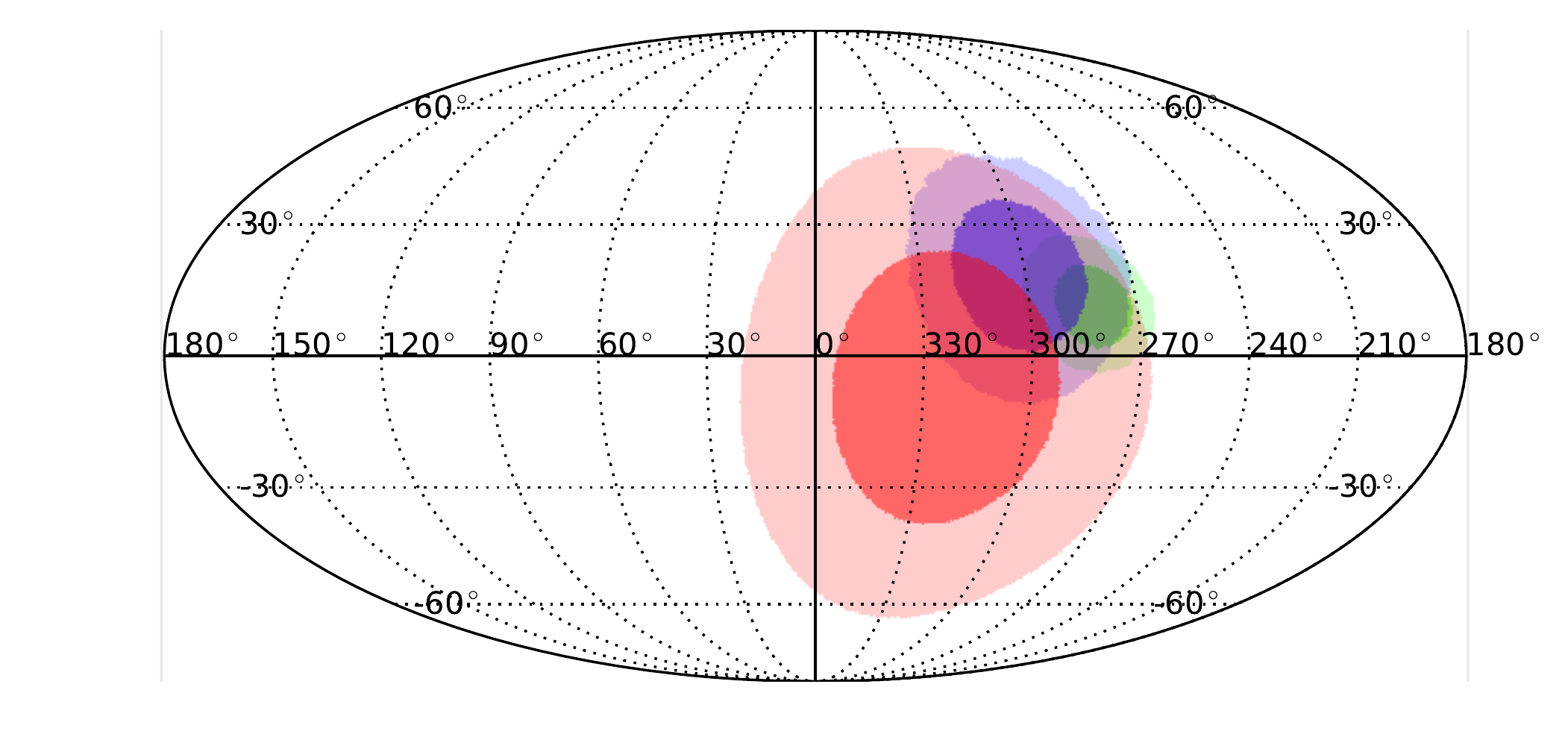}
  \caption{\label{fig:compare_bulk_flows} We have drawn the 68\% and 95\% probability limit for the kSZ signal direction and some of the bulk flows given in Table~\ref{tab:bulk_flows}. Green: \protect\cite{WFH09}, limited to 20\Mpch{} Gaussian. Blue: \protect\cite{LTMC10}, limited to 50\Mpch{} Top-hat. Red: kSZ signal, template generated with galaxies $cz \le 5 000$~\kms{}.}
\end{figure}

When only galaxies within 50\Mpch{} are used in the kSZ template, we find $|V|=533\pm 263$~\kms{} (after correction for ``error biasing''\footnote{The velocity amplitude is a positive number so random errors will bias the true amplitude upward. We correct for this by subtracting off in quadrature the uncertainties in each Cartesian component of the flow.}) in the direction $l=324\pm 27$, $b=-7 \pm 17$.  This is similar in direction and amplitude to recent bulk flows based on classical distance measurements. For example, at a similar depth, the ``Seven Samurai'' (7S) found $V = 599 \pm 104$ \kms{} in the direction $l = 312\pm11$, $b = 6 \pm 10$ \citep{DreLynBur87} using the $D_n$ -- $\sigma$ distance indicator for early-type galaxies. Table~\ref{tab:bulk_flows} shows three recent bulk flow measurements derived using three different methods, along with the characteristic scale for averaging the peculiar velocities. We have selected these three bulk flows because they probe similar volumes to that probed by the kSZ measurement of Line 1 of Table~\ref{tab:results}. For example, the ``SHALLOW'' sample analysed by \cite{WFH09} includes measurements from surface brightness fluctuations \citep{TonDreBla01} and the ENEAR sample \citep{BerAlodaC02}, an updated version of the 7S sample quoted above, whereas the SFI++ sample contains the latest Tully-Fisher distances. The result from \cite{LTMC10}  is instead is a predicted bulk flow, assuming that 2MASS-selected galaxies trace the mass. Figure \ref{fig:compare_bulk_flows} shows the direction of the bulk flow for kSZ measurement (limited to 50 \Mpch{}) and for the data sets discussed above.
The directions are in good agreement.

\begin{table*}
  \caption{\label{tab:bulk_flows} Summary of bulk flows from in the literature with similar depths as the kSZ measurement.}
  \begin{center}
    \begin{tabular}{ccccccccc}
      \hline
      Reference & Scale & Filter & $V_x$ & $V_y$ & $V_z$ & $|V|$ & $l$ & $b$ \\
                & (\Mpch) &      & (\kms) & (\kms) & (\kms) & (\kms) & ($^\circ$) & ($^\circ$) \\
      \hline
      \cite{WFH09} & 20  & Gaussian & $85 \pm 27$ & $-228 \pm 43$ & $53 \pm 26$ & $253 \pm 40$ & $291 \pm 7$ & $12 \pm 6$ \\
                   & SHALLOW  & MLE & $190 \pm 33$ & $-259 \pm 29$ & $110 \pm 23$ & $342 \pm 30$ & $306 \pm 6$ & $19 \pm 4$ \\
                   & SFI++    & MLE & $117 \pm 28$ & $-331 \pm 28$ & $96 \pm 20$ & $365 \pm 27$ & $290 \pm 5$ & $15 \pm 3$ \\
      \cite{LTMC10} & 50 & Top-hat  & $227 \pm 77$ & $-270 \pm 78$ & $136 \pm 78$ & $393 \pm 75$ & $310 \pm 13$ & $20 \pm 11$ \\
      \hline
    \end{tabular}
  \end{center}
  \parbox{\hsize}{
  Note: In the case of ``SHALLOW'' and ``SFI++'', we have used the Maximum Likelihood Estimator (MLE) values quoted in \cite{WFH09}. The value given reported as 20\Mpch{} Gaussian is obtained using the minimum variance estimator applied on the ``COMPOSITE'' sample.
  }
\end{table*}

Given the agreement in direction, we may assume that the amplitude itself is significant. The amplitude of the kSZ signal, expressed as bulk flow components $\mathbf{V}_\text{bulk,kSZ}$, may be modelled as the following:
\begin{equation}
	\mathbf{V}_{\text{bulk,kSZ}} = \tilde{f} \mathbf{V}_{\text{bulk}},
\end{equation}
with $\tilde{f}=f_e (\sigma / 160\text{ km s}^{-1})^2$, $f_e$ being the fraction of free electrons and $\sigma$ modelling the unknown normalisation of the density profile of the galactic plasma halo. $\tilde{f}$ is taken as a free parameter for this subsection, $\mathbf{V}_\text{bulk}$ the bulk flow components as determined by other methods listed in Table~\ref{tab:bulk_flows}. We may derive the posterior probability of $\tilde{f}$ using Bayes identity, which, up to a normalisation, is:
\begin{multline}
	P(\tilde{f}|\mathbf{V}_{\text{bulk,kSZ}},\mathbf{V}_{\text{bulk}}) \propto \\
	  \int \mathrm{d}^3 \mathbf{V}_\text{bulk,true} P(\mathbf{V}_{\text{bulk,kSZ}} | \tilde{f}, \mathbf{V}_{\text{bulk, true}}) P(\mathbf{V}_{\text{bulk, true}}), \label{eq:posterior_fe}
\end{multline}
with 
\begin{equation}
	P(\mathbf{V}_{\text{bulk,kSZ}} | \tilde{f}, \mathbf{V}_{\text{bulk}}) = \mathcal{G}(\tilde{f} \mathbf{V}_{\text{bulk,true}}, \mathbf{\sigma}_\mathrm{kSZ}),
\end{equation}
and 
\begin{equation}
	P(\mathbf{V}_{\text{bulk,true}}) = \mathcal{G}(\mathbf{V}_{\text{bulk}}, \mathbf{\sigma}_\mathrm{bulk}).
\end{equation}
In the above expressions, we have used the notation $\mathcal{G}(\mathbf{m},\mathbf{s})$ for a three dimensional Gaussian probability, with the mean $m_i$ and the standard deviation $s_i$ for the same component. $\mathbf{\sigma}_\mathrm{kSZ}$ ($\mathbf{\sigma}_\mathrm{bulk}$ respectively) is given by the errors on the components quoted in Table~\ref{tab:results} (Table~\ref{tab:bulk_flows} respectively).
Effectively the posterior probability \eqref{eq:posterior_fe} may be written as a likelihood function
\begin{multline}
	\mathcal{L}(\tilde{f}) = \sum_{i=1}^3 \left[\left( \frac{V_{i,\text{bulk,kSZ}}-\tilde{f} V_{i,\text{bulk}}}{\tilde{f}^2 \sigma^2_{i,\text{bulk}} +\sigma^2_{i,\text{kSZ}}}\right)^2 \right.\\ 
	\left.	+ \log \left(\tilde{f}^2 \sigma^2_{i,\text{bulk}} +\sigma^2_{i,\text{kSZ}}\right)\right]. \label{eq:fe_likelihood}
\end{multline}

\begin{figure}
	\includegraphics[width=\hsize]{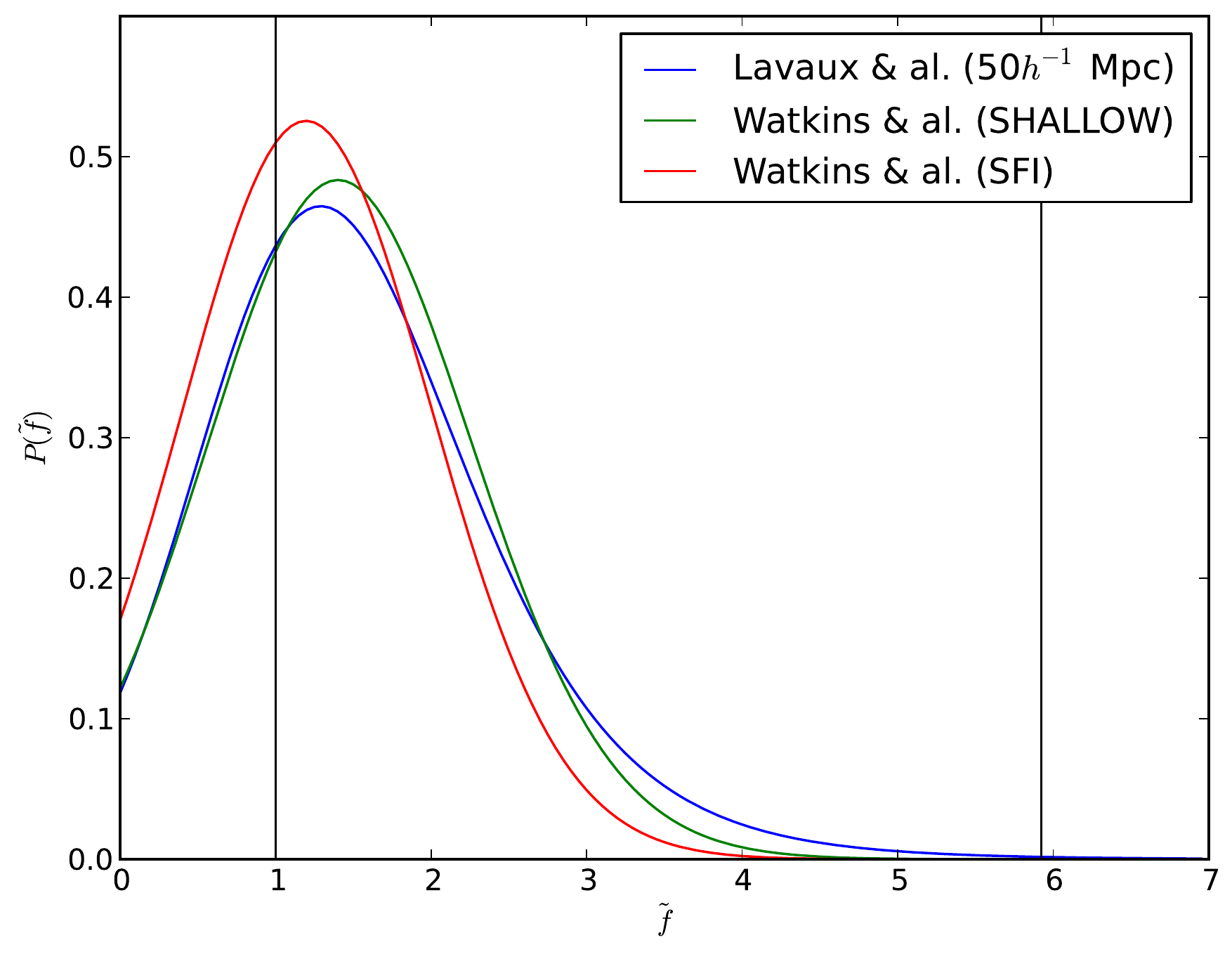}
	\caption{\label{fig:bfrac_posterior} The posterior distribution for the $\tilde{f}$ factor assuming different measured bulk flows for a single kSZ measurement given by line 1.  The left (right respectively) black solid vertical line correspond to $\tilde{f}=1$ ($\tilde{f}=\Omega_m/\Omega_b=5.92$ respectively).}
\end{figure}

We show the posterior distribution of the parameter $\tilde{f}$ in Figure~\ref{fig:bfrac_posterior} for different choices of the published bulk flows against one single of our kSZ measurement. We are taking the line 1 of Table~\ref{tab:results} as the fiducial kSZ measurement. Assuming the bulk flow for the SHALLOW sample in \cite{WFH09}, the posterior peaks at $\tilde{f} = 1.4^{+1.02}_{-0.99}$ (68\% confidence limit). For the bulk flow of the SFI sample, this value becomes $\tilde{f}=1.2^{+0.95}_{-0.94}$. Assuming \cite{LTMC10} cut at 50\Mpch{} top-hat, the posterior peaks at $\tilde{f}=1.29^{+1.1}_{-0.92}$. For information, we have shown two extreme choices of choosing the amount of baryons in galaxies. The choice $\tilde{f}=1$ corresponds to considering that galaxies have a dark matter halo and all baryons are ionised in the plasma halo. For indicative purposes, we have shown $\tilde{f}=\Omega_\text{m,fid}/\Omega_\text{b}=5.92$ in Figure~\ref{fig:bfrac_posterior}, $\Omega_\text{m,fid}=0.258$. We would measure such a value if the density of baryons were equal to the mean density of dark matter. 
We note that all posteriors are in agreement with a choice of $\tilde{f}=1$. The ratio of the posterior at its maximum to its value for $\tilde{f}=0$ in the best case (green curve, SHALLOW sample) is $3.97$. For the blue and red curve, this ratio is $3.93$ and $3.08$ respectively. According to \cite{HJef}, a ratio of $3.97$ corresponds to \emph{substantial evidence} against the null hypothesis there are no free electrons in galaxy halos .

\section{Discussion and Future Work}
\label{sec:discuss}

There are several ways in which this measurement can be improved in the near future. First, improved CMB data is anticipated from SPT \citep{SPT_pub11} and from the  {\sc Planck} mission.  We discuss below the improvement in signal-to-noise expected from  {\sc Planck}.  Moreover, we anticipate a more detailed treatment of the peculiar velocity field, an improved understanding of the halo model and of the distribution of free electrons around galaxies.

\subsection{Forecast for {\sc Planck}}
\label{sec:tests}

In this section, we estimate the noise covariance matrix of the three components of the bulk flows for the maps obtained from the {\sc Planck} probe. The inverse covariance matrix is given by Equation~\eqref{eq:covariance_A}. Neglecting the impact of the other foreground components, assuming the noise is homogeneous on the sky and no masking, the covariance matrix $C_B$ of the kSZ bulk flow becomes diagonal in spherical harmonic space and simplifies to:
\begin{multline}
  C_{B,i,j}^{-1} = t_i^\dagger \left( \sum_{c,c'} B_c C^{-1}_{c,c'} B_c' \right) t_j = \\
  \sum_\ell (2\ell+1) T_{i,j,\ell} \\
  \sum_{c,c'} B_c \left(\delta_{c,c'} N^{-1}_{c,\ell} -  \frac{B_{c,\ell} B_{c',\ell} C_\ell N^{-1}_{c,\ell} N^{-1}_{c',\ell}}{1 + C_\ell \sum_c N^{-1}_{c,\ell} B_{c,\ell}^2}\right)B_{c'},
\end{multline}
with $T_{\ell,i,j}$ the angular cross power spectrum of the kSZ template between the component $i$ and $j$ of the bulk flow. 
For a single channel, this reduces to:
\begin{equation}
  C_{B,i,j}^{-1} = \sum_\ell (2\ell+1) \frac{T_{i,j,\ell} B_{\ell}^2}{N_\ell + C_\ell B_{\ell}^2}.
\end{equation} 
We give the forecast estimate for the channel W of WMAP7 in Table~\ref{tab:forecast_vs_obtained}. In this same table, we also report the actual errors that we obtained using the full method. To make a forecast for the sensitivity that we expect using {\sc Planck} maps, we consider the map at a frequency of 100~GHz. According to \cite{Planck}, the noise at resolution of $N_\text{side}=1024$ is $\sim 7$~$\mu$K for the temperature. The beam is modelled using a Gaussian with a Full-Width-Half-Maximum equal to $10$~arcmins. The last column of the Table~\ref{tab:forecast_vs_obtained} gives the standard deviation expected for the velocity components of the bulk flow with these parameters. We have also tried our forecast with a more complex model based on the published Planck noise properties \protect\citep{Planck_V}, which increases the errors reported in  Table \ref{tab:forecast_vs_obtained}  by $\sim$20\%. Also note that these are the errors for the 50\Mpch{} cut. {\sc Planck} is able to reach much more distant galaxies. 

There is a good agreement between the estimated value of the noise for WMAP data and the actual noise amplitude that we measure using the complete analysis of Section~\ref{sec:stat_method}. This gives us some confidence that the projection for {\sc Planck} is realistic. We note that, mainly due to the reduction of the beam width, the noise should go down by a factor of two to three by using {\sc Planck} data compared to WMAP data. This estimate does not take into account possible additional foreground contamination. If the bulk flows stay at the same amplitude as we have estimated from WMAP data, that could give us a $\sim 5.4\sigma$ detection of the kSZ effect. If the bulk flow is at the level of that found for the   SFI++ sample  \citep{WFH09}, the significance could decrease to $2.2\sigma$. It must be nonetheless noted that we have neglected all contributions from the multi-channel analysis in this prediction. The high-frequency channels would clearly help to  remove the contribution of the tSZ signal. Additionally, the temperature data from the {\sc Planck} mission have sufficient angular resolution to probe deeper flows on scales up to 200\Mpch{} scale with better sensitivity. As a final note, we assume that the kSZ signal has the same amplitude as what is expected from the bulk flow given by the SHALLOW sample in Table~\ref{tab:bulk_flows}. In this case, we forecast that the constraints on the baryon density in plasma haloes of galaxies from the {\sc Planck} mission should be $\Delta \tilde{f} \sim 0.2$, instead of $\sim 0.9$ as obtained in Section~\ref{sec:comparison_bf}. 

\begin{table}
  \begin{center}
    \begin{tabular}{cccc}
      \hline
      Component & WMAP7  & WMAP7  & PLANCK  \\
                & obtained & forecast & forecast \\
                & (\kms) & (\kms) & (\kms) \\
      \hline
      x & 290 & 305 & 120 \\
      y & 285 & 277 & 111 \\
      z & 200 & 224 & 93 \\
      \hline
    \end{tabular}
  \end{center}
  \caption{\label{tab:forecast_vs_obtained} Standard deviation estimates on bulk flow components for the kSZ templates generated from the 2M++ catalogue cut at a depth 50\Mpch{}.}
\end{table}

\subsection{Model improvements}

There are several ways in which the model for the free electron momentum field will be improved.
\begin{enumerate}

\item The model for the flow field will be updated. In this paper, for simplicity, we have modelled the peculiar velocity field as a simple bulk flow, a low order moment of the peculiar velocity field. However, as suggested by \cite{HDS09} and \cite{SZLJP11}, by using linear perturbation theory and an assumption about the relationship between mass and light in haloes (linear biasing or the halo model), we can predict the peculiar velocities of individual galaxies \citep{YahSanTam80, StrDavYah92, Hud94b} to an accuracy of $\sim 150$ \kms{} \citep{WilStr98} or better using a non-linear reconstruction \citep{LMCTBS08, LTMC10}.  These predictions scale with the parameter $\beta = f(\Omega_m)/b$ where $f \sim \Omega_m^{0.55}$ for $\Lambda$CDM and $b$ is a bias factor, and so the kSZ effect will allow us to get an independent measurement of this parameter.

\item  We are using galaxies as a tracer of the distribution of free electrons, but we can improve this model in several ways. First, the scaling of halo mass and radius with galaxy luminosity or stellar mass can be updated using results from abundance matching \citep{MarHud02, YanMovan03, GuoWhiLi10} or weak lensing studies of low redshift galaxies \citep{ManSelKau06,VelvanHoe12}.  Second, high-resolution hydrodynamical cosmological simulations can be used to get a better handle on the link between galaxy halo profiles and their cross-correlations with the distribution of free electrons \citep[as in e.g.][]{SZLJP11, PJSP12}.

\end{enumerate}


\section{Conclusion}
\label{sec:conclusion}

We have developed and applied a method for estimating bulk flows from kSZ signals in temperature maps of the Cosmic Microwave Background. This method is based on a likelihood analysis of the amplitude of a set of template maps derived from catalogues of galaxies. Assuming an isothermal model for the galaxy plasma halo, we have derived template maps for the kinetic Sunyaev-Zel'dovich from the 2M++ galaxy catalogue. The thermal Sunyaev-Zel'dovich signal is derived in a similar fashion from the RBC galaxy cluster catalogue, assuming  the model described in \cite{ALS05} for the density and temperature of the gas. To estimate the extent of the contamination by point sources residing in galaxies, we have also generated a set of templates from the 2MASS-XSC. We have fitted, independently in each channel, the standard WMAP7 foreground templates, the monopole and the dipole. 

The amplitude of the tSZ signal is measured with a high accuracy at $13\sigma$ confidence. We have found evidence at the  90\% confidence level for the presence of a kSZ signal in WMAP7 data. We have not detected any significant contamination by point sources. All the results are presented after having marginalised over the Galactic foreground emissions. 

Interpreting the kSZ signal as the signature of a bulk flow, we derive an amplitude, when only galaxies within 50\Mpch{} are included in the template, of $|V|=533\pm 263$~\kms{}, in the direction $l=324\pm 27$, $b=-7 \pm 17$, in excellent agreement with the "Seven Samurai" bulk flow \citep{DreLynBur87} and other recent determinations on similar scales. If we extend selection up to 200\Mpch{}, we find $|V|=284\pm 187$~\kms{}, in the direction $l=26\pm 35$, $b=-17 \pm 19$. The kSZ bulk flow at 50\Mpch{} is statistically in agreement with other measurements obtained from either galaxy peculiar velocities, observed or modelled. This comparison can be translated into a measurement of the fraction of baryon in galaxies. We found substantial evidence of baryons in the galaxy plasma halo, compatible with the mean density of baryons derived from WMAP7 \citep{WMAP7_Cosmo}. {\sc Planck} data should be able to reduce the statistical error bars on the kSZ measurement by a factor two to three. If the signal stays at the present value, the evidence may be increased to $5.4\sigma$, excluding improvements due to the high frequency channels for removing tSZ contamination. Realistically, if the bulk flow is as given by \cite{WFH09}, then the evidence would be at the $2.2\sigma$ level. 

\section*{Acknowledgements}

NA and MJH acknowledge the preliminary work of Matthias M\"uller on a similar project.

This publication makes use of data products from the Two Micron All Sky Survey, which is a joint project of the University of Massachusetts and the Infrared Processing and Analysis Center/California Institute of Technology, funded by the National Aeronautics and Space Administration and the National Science Foundation. 

GL acknowledges support from CITA National Fellowship and financial support from the Government of Canada Post-Doctoral Research Fellowship and NSF Grant AST 07-08849, AST 09-08693.  MH acknowledges support from NSERC. 

Research at Perimeter Institute is supported by the Government of Canada through Industry Canada and by the Province of Ontario through the Ministry of Research and Innovation.

This work was made possible by the facilities of the Shared Hierarchical Academic Research Computing Network (SHARCNET:\url{http://www.sharcnet.ca}) and Compute/Calcul Canada.

We acknowledge the use of the Legacy Archive for Microwave Background Data Analysis (LAMBDA). Support for LAMBDA is provided by the NASA Office of Space Science.

Some of the results in this paper have been derived using the HEALPix\footnote{\url{http://healpix.jpl.nasa.gov}} \citep{HEALPIX} package.


\appendix
\onecolumn

\section{Derivation of the likelihood from the model}
\label{app:likelihood}

In this Appendix, we derive the Equation~\eqref{eq:chi2} directly from the model assumptions. We start by expressing that the primary CMB fluctuations and the instrumental noise by channel are Gaussian. They have known covariance matrices $S$, for the primary fluctuations, and $N_c$, for the channel $c$ of the instrument. $N_c$ may include masking effect. $\tilde{N}_c$, as defined in Equation~\eqref{eq:mask_noise} or $N_c$ may both be used in the following. The likelihood of the templates, assuming the amplitudes $\alpha_{i,\rho}$ and the modes $s$ of the primary fluctuations and given the input data, is thus:
\begin{equation}
  \chi^2(\mathbfit{s},\{\alpha_{i,\rho}\}|\{\mathbfit{d}_c\}) = 
    \mathbfit{s}^\dagger \mathbf{S}^{-1} \mathbfit{s} + 
        \sum_{c} \mathbfit{r}^\dagger_c(\mathbfit{s},\mathbfit{d}_c,\{\alpha_{i,\rho}\}) \mathbf{N}_c^{-1} \mathbfit{r}_c(\mathbfit{s},\mathbfit{d}_c,\{\alpha_{i,\rho}\}),
\end{equation}
with
\begin{equation}
  \mathbfit{r}_c = \mathbfit{d}_c - \mathbfit{B}_c \left(\mathbfit{s} + \sum_i \alpha_{i,\sigma_i(c)} \mathbfit{t}_{i,c}\right),
\end{equation}
the residual. The above terms can be reordered such that $s$ appears only in one quadratic term:
\begin{equation}
  \chi^2(\mathbfit{s},\{\alpha_{i,\rho}\}|\{\mathbfit{d}_c\}) =
    (\mathbfit{s} - \mathbfit{s}_0)^\dagger \mathbf{S}^{-1/2} \mathbf{D} \mathbf{S}^{-1/2} (\mathbfit{s}-\mathbfit{s}_0) 
    - \mathbfit{s}^\dagger_0 \mathbf{S}^{1/2} \mathbf{D}^{-1} \mathbf{S}^{1/2} \mathbfit{s}_0
    + \sum_{c} \mathbfit{d}^\dagger_c \mathbf{N}_c^{-1} \mathbfit{d}_c, 
\end{equation}
with
\begin{equation}
  \mathbfit{s}_0 = \sum_c \mathbf{S}^{1/2} \mathbf{D}^{-1} \mathbf{N}_c^{-1} \mathbf{B}_c \left(\mathbfit{d}_c - \sum_i \alpha_{i,\sigma_i(c)} \mathbfit{t}_{i,c}\right),
\end{equation}
and $\mathbf{D}$ as defined in Equation~\eqref{eq:D}. The likelihood is now split into two parts:
\begin{equation}
   \chi^2(\mathbfit{s},\{\alpha_{i,\rho}\}|\{\mathbfit{d}_c\}) = (\mathbfit{s} - \mathbfit{s}_0)^\dagger \mathbf{S}^{-1/2} \mathbf{D} \mathbf{S}^{-1/2} (\mathbfit{s}-\mathbfit{s}_0) + 
     \sum_{c,c'} \left(\mathbfit{d}_c - \sum_i \alpha_{i,\sigma_i(c)} \mathbfit{t}_{i,c}\right) \mathbf{C}^{-1}_{c,c'} \left(\mathbfit{d}_{c'} - \sum_j \alpha_{j,\sigma_j(c')} \mathbfit{t}_{j,c'}\right).
\end{equation}
The marginalisation according to primary CMB fluctuations consists in dropping the first part, which is the only one depending on $s$. Equation~\eqref{eq:chi2} immediately follows.

\section{Accelerating the precomputation of the likelihood}
\label{app:precomputation}

For each template $\mathbfit{t}_{k,c}$ of the signal $k$ in the frequency channel $c$, we define a weighted template
\begin{equation}
  \tilde{\mathbfit{t}}_{k,c} = \mathbf{S}^{1/2} \mathbf{D}^{-1} \mathbf{S}^{1/2} \mathbf{B}_{c'} \tilde{\mathbf{N}}^{-1}_{c'} \mathbfit{t}_{k,c}.
\end{equation}
The Equations \eqref{eq:maxlikelihood_alpha} and \eqref{eq:covariance_A} may be rewritten in terms of uniquely these weighted templates. We expand the Equation~\eqref{eq:optimal_weighting} in the Equations~\eqref{eq:maxlikelihood_alpha} and \eqref{eq:covariance_A}. After some algebra, we find the expression, numerically simpler, of the covariance matrix 
\begin{equation}
  \mathcal{\mathbf{A}}_{(i,\rho),(j,\nu)} =  
      \sum_{\substack{c,c' \\ \sigma_i(c)=\rho, \sigma_j(c')=\nu}} \left( \mathbfit{t}^{\dagger}_{i,c} \mathbf{B}_c \tilde{\mathbf{N}}_c^{-1} \mathbf{B}_c' \mathbfit{t}_{j,c'} - \mathbfit{t}^{\dagger}_{i,c} \mathbf{B}_c \tilde{\mathbfit{t}}_{j,c'} \right),
\end{equation}
and for the maximum likelihood estimate
\begin{equation}
  \alpha = \sum_{j,\nu} \mathcal{\mathbf{A}}_{(i,\rho),(j,\nu)}^{-1} \left(\sum_{\substack{c \\ \sigma_j(c)=\nu}} \sum_{c'} \tilde{\mathbfit{t}}_{j,c}^\dagger \mathbfit{d}_{c'}\right).
\end{equation}
Only matrices that are diagonal either in pixel space or in harmonic space are involved in the two above equations. We reduce the computational time complexity from $\mathcal{O}(N_t^2 \times N_c^2)$ to $\mathcal{O}(N_t \times N_c)$ in terms of the inversion $\mathbf{C}^{-1}_{c,c'}$.

\section{Normalisation of the galaxy density profiles}
\label{app:isothermal_sphere}

The mass of a galaxy, for which the profile is limited to the radius $R=m_V R_V$, $R_V$ being the virial radius defined in \cite{FP06}, is:
\begin{equation}
  M(m_V) = \frac{2 \sigma^3}{\sqrt{50} G H_0} m_V
\end{equation}
Assuming an homogeneous distribution of galaxies, with a number of density $n_*$, and that the total amount of dark matter is traced by galaxies, the density of dark matter is 
\begin{equation}
  \rho_{m,g} = M(m_v) n_* = \frac{2 n_* \sigma^3}{\sqrt{50} G H_0} m_V.
\end{equation}
Comparing this quantity to the measured mean matter density,
\begin{equation}
  \rho_m = \frac{3 H_0^2}{8\pi G} \Omega_\text{m},
\end{equation}
gives
\begin{equation}
  \frac{\rho_{m,g}}{\rho_m} = (0.41\pm 0.04)\, m_V,
\end{equation}
with $n_*=(1.13 \pm 0.02)\,10^{-2} h^3$~Mpc$^{-3}$ \citep{LH11}, $\Omega_\text{m}=0.266\pm 0.028$ (WMAP7 alone) and $\sigma=160$~\kms{} \citep{FP06}. The error bar were added in quadrature to obtain the error on the fraction. This result motivates the choice of the effective radius for the matter halo around galaxies. Taking $m_V=2.4$ yields to an agreement between the matter density estimated using galaxies and the one obtained from cosmological probes. 

We have tried to use a Navarro-Frenk-White white profile instead of the isothermal profile. To normalise it in the same way, we have enforced that the mass within a virial radius of the isothermal profile is the same as for the isothermal profile. 

\section{Statistics of the KSZ estimator of the bulk flow}
\label{app:ksz_window}

For the purpose of understanding the scales probed by the estimator of Equation~\eqref{eq:maxlikelihood_alpha} in the case of the kSZ effect, we model the full observed temperature fluctuations due to the kSZ signal like:
\begin{equation}
  d_\mathrm{obs}(\hat{n}) = A_{\mathrm{kSZ}} \int_0^\infty \mathrm{d}r\; n^{\mathrm{r}}_e(r\hat{n}) \mathbfit{v}(r\hat{n}) \cdot \hat{n},
\end{equation}
with $v^a(\mathbfit{r})$ the $a$ component of the velocity field taken at position $\mathbfit{r}$, $A_\mathrm{kSZ}$ the normalisation of the kinetic Sunyaev-Zel'dovich, $n^{\mathrm{r}}_e(\mathbfit{r})$ the density of electrons at position $\mathbfit{r}$ and $\hat{n}$ the direction we are looking in the sky. In this work, we extract the bulk flow component by fitting a modulated dipole on the projected electron distribution. Thus the model is:
\begin{equation}
  d_\mathrm{m}(\mathbfit{V}, \hat{n}) = A_\mathrm{kSZ} \int_0^{R_S} \mathrm{d}r\; n^{\mathrm{m}}_e(r\hat{n}) \mathbfit{V} \cdot \hat{n},
\end{equation}
with $R_s$ the limiting size of the survey used to fit the velocity dipole, and $n^{\mathrm{m}}_e(\mathbfit{r})$ is the modelled density of electrons.

We may write the fitting procedure in spherical harmonic space. In this space, it is possible to marginalise analytically according to CMB fluctuation and to take into account beaming effects. The $\chi^2$ corresponding to the likelihood to maximise may be written as:
\begin{equation}
  \chi^2(\mathbfit{V}) = \sum_{\ell, m} \frac{1}{C_\ell} \left|a_{\ell m}^{\mathrm{obs}} - a_{\ell, m}^{\mathrm{m}}(\mathbfit{V})\right|^2,
\end{equation}
with $C_\ell = C_\ell^\text{CMB} + B_\ell^{-2} N_\ell$,  $N_\ell$ being the noise power spectrum of the instrument, $B_\ell$ the beam of the instrument, $a^\text{obs}_{\ell m}$ the spherical harmonic coefficient of the temperature data $d_\mathrm{obs}(\hat{n})$
\begin{equation}
  a_{\ell m}^{\mathrm{obs}} = \int \textrm{d}^2\hat{n}\; Y^*_{\ell,m}(\hat{n}) d_\text{obs}(\hat{n}),
\end{equation}
and the similar relation between $a_{\ell m}^{\mathrm{m}}(\mathbfit{V})$ and $d_\text{m}(\hat{n},\mathbfit{V})$. By maximising the likelihood of the bulk flow component, thus minimising the above $\chi^2$ according to the components $V^a$, we obtain the following algebraic expression:
\begin{equation}
  V^a = \sum_{b=1}^3 (\mathcal{M}^{-1})^{a,b} \mathcal{V}^b  \label{eq:bf_estimator}
\end{equation}
with
\begin{align}
  \mathcal{M}^{a,b} & = 2 \sum_{\ell,m} \frac{1}{C_\ell} \left(\int \mathrm{d}^2\hat{r}\; \int_{r=0}^{R_S} \text{d}r\; Y^*_{\ell,m}(\hat{r}) n_e(r\hat{r}) \hat{r}^a\right) \left(\int \mathrm{d}^2\hat{s}\; \int_{s=0}^{R_S}\text{d}s\; Y_{\ell,m}(\hat{s}) n_e(s\hat{s}) \hat{s}^b \right), \\
  \mathcal{V}^a & = 2 \sum_{\ell m} \frac{1}{C_\ell} \int \mathrm{d}^2\hat{r} \mathrm{d}^2\hat{s} \int_{r=0}^{R_S}\int_{s=0}^{+\infty} \mathrm{d}r\mathrm{d}s\;  Y^*_{\ell,m}(\hat{r}) Y_{\ell,m}(\hat{s}) n_e(r\hat{r}) n_e(s\hat{s}) \hat{r}^a \hat{s}^b v^b(s\hat{s}).
\end{align}
We look for the average properties of the estimator given in Eq.~\eqref{eq:bf_estimator} of the bulk flow $\mathbfit{V}$. To do this, we want to compute the average response of the estimator for different velocity fields and for a fixed positions of tracers. Thus the densities $n_e$ are kept constant, but the velocity field $\mathbfit{v}(\mathbfit{r})$ is left free. A useful tool to quantify the averaging properties is to consider the covariance $\langle V^a V^b \rangle$. Through $\mathcal{M}^{a,b}$, it is directly related to $\langle \mathcal{V}^a \mathcal{V}^b \rangle$:
\begin{multline}
  \langle \mathcal{V}^a \mathcal{V}^b \rangle = \\ 4 \sum_{\ell,m} \sum_{\ell',m'} \frac{1}{C_\ell C_{\ell'}} \int \mathrm{d}^2\hat{r} \mathrm{d}^2\hat{s} \mathrm{d}^2\hat{r'} \mathrm{d}^2\hat{s'} \int_{r=0}^{R_S}\int_{s=0}^{+\infty}\int_{r'=0}^{R_S}\int_{s'=0}^{+\infty} \mathrm{d}r\mathrm{d}s\; Y^*_{\ell,m}(\hat{r}) Y_{\ell,m}(\hat{s}) Y^*_{\ell',m'}(\hat{r}') Y_{\ell',m'}(\hat{s}') \\
  \times n_e(\mathbfit{r}) n_e(\mathbfit{s})  n_e(\mathbfit{r}') n_e(\mathbfit{s}') \hat{r}^a \hat{r'}^{b}  \hat{s}^c \hat{s'}^{d} \langle v^c(\mathbfit{s}) v^d(\mathbfit{s}') \rangle. \label{eq:VV_covariance}
\end{multline}
Assuming that the linear perturbation theory is a good description of the velocity field, that the statistics of density fluctuations is purely Gaussian and described by a power spectrum $P(k)$, then the covariance of the velocity field is
\begin{equation}
  \langle v^c(\mathbfit{s}) v^d(\mathbfit{s}') \rangle = \frac{a^2 H^2 f^2}{2\pi^2}\int_{k=0}^{+\infty}\mathrm{d}k\; P(k) \int \mathrm{d}^2\hat{k}\; \hat{k}^c \hat{k}^d \mathrm{e}^{ik \hat{k}\cdot (\mathbfit{s}-\mathbfit{s}')},
\end{equation}
with $f=\mathrm{d}\log D/\mathrm{d}\log a$, $D$ the growth factor, $a$ the scale factor of the Universe and $H$ the Hubble constant.
We may now rewrite Eq.~\eqref{eq:VV_covariance} 
\begin{equation}
  \langle \mathcal{V}^a \mathcal{V}^b \rangle = a^2 H^2 f^2 \int_{k=0} \frac{\mathrm{d} k}{6\pi^2} P(k) \widetilde{\mathcal{W}}^{a,b}(k), \label{eq:covariance_pspectrum}
\end{equation}
with
\begin{multline}
  \widetilde{\mathcal{W}}^{a,b}(k) =  4 \sum_{\ell,m} \sum_{\ell',m'} \frac{1}{C_\ell C_{\ell'}} \int \mathrm{d}^2\hat{r} \mathrm{d}^2\hat{s} \mathrm{d}^2\hat{r'} \mathrm{d}^2\hat{s'} \int_{r=0}^{R_S}\int_{s=0}^{+\infty}\int_{r'=0}^{R_S}\int_{s'=0}^{+\infty} \mathrm{d}r\mathrm{d}s\;  Y^*_{\ell,m}(\hat{r}) Y_{\ell,m}(\hat{s}) Y^*_{\ell',m'}(\hat{r}') Y_{\ell',m'}(\hat{s}') \\
  \times n_e(\mathbfit{r}) n_e(\mathbfit{s})  n_e(\mathbfit{r}') n_e(\mathbfit{s}') \hat{r}^a \hat{r'}^{b}  \hat{s}^c \hat{s'}^{d} \int\frac{3 \mathrm{d}^2\hat{k}}{4\pi} \hat{k}^c \hat{k}^d \mathrm{e}^{i \mathbfit{k}\cdot (\mathbfit{s}-\mathbfit{s'})}. 
\end{multline}
Up to a multiplication by the inverse of $\mathcal{M}^{a,b}$, we call this function the window function of the estimator. According to Eq.~\eqref{eq:covariance_pspectrum}, this function is a transfer function to apply on the power spectrum of density fluctuations to obtain the covariance of the estimated bulk flow. $\widetilde{\mathcal{W}}^{a,b}$ indicates what are the scales probed by the estimator. Other similar window functions have been derived in the literature in the context of survey of peculiar velocities of galaxies \citep{WFH09}. 
Some factorisation may be done:
\begin{equation}
  \widetilde{\mathcal{W}}^{a,b}(k) = 4 \sum_{\ell,m} \sum_{\ell',m'} \frac{1}{C_\ell C_{\ell'}} g_{\ell,m} g_{\ell',m'} \int_{s=0}^{+\infty}  \int_{s'=0}^{+\infty} Y_{\ell,m}(\hat{s}) Y_{\ell',m'}(\hat{s}') n_e(\mathbfit{s})  n_e(\mathbfit{s}') \hat{s}^c \hat{s'}^{d} \int\frac{3 \mathrm{d}^2\hat{k}}{4\pi} \hat{k}^c \hat{k}^d \mathrm{e}^{i \mathbfit{k}\cdot (\mathbfit{s}-\mathbfit{s'})}. \label{eq:W_attempt0}
\end{equation}
with
\begin{equation}
  g^a_{\ell,m} = \int \mathrm{d}^2\hat{n} \int_{r=0}^{R_S}\text{d}r\; \hat{n}^a Y^*_{\ell,m}(\hat{n}) n_e(\mathbfit{r}).
\end{equation}
$g^a_{\ell,m}$ is exactly the spherical harmonic representation of the kSZ templates that we are using to do the fitting of the component $a$ of the bulk flow in Section~\ref{sec:ksz_template}. The part after the integration symbol in the Equation~\ref{eq:W_attempt0} corresponds to the complete kSZ signal generated by the whole observable universe.

\begin{figure}
  \begin{center}
    \includegraphics[width=.6\hsize]{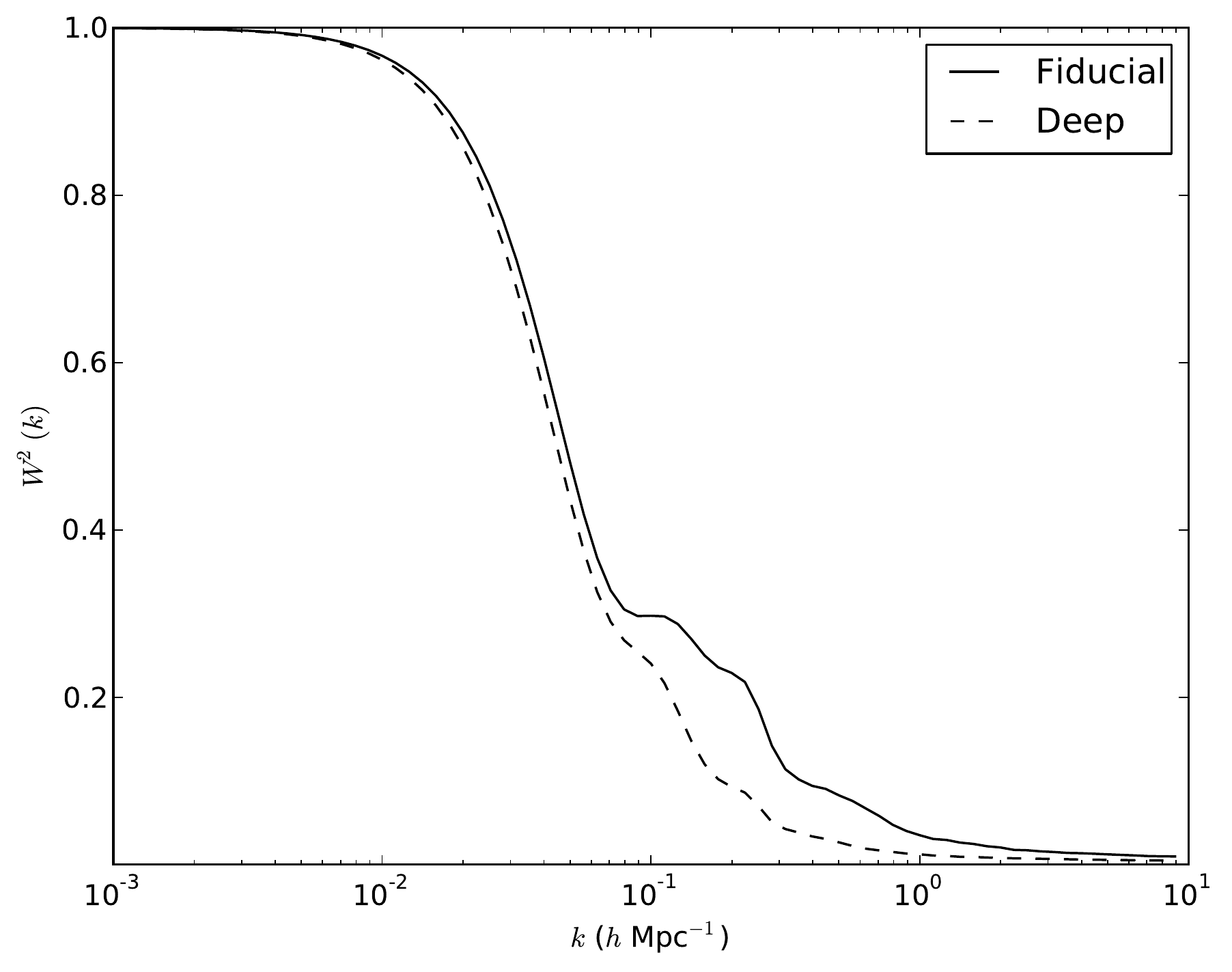}
  \end{center}
  \caption{\label{fig:test_convergence} We show the problem of the convergence of the window function for  $\mathcal{W}^{x,x}(k)$. The fiducial curve corresponds to taking the same sample for the galaxies $\{\mathbfit{x}_i\}$ in Equation~\eqref{eq:window_prefinal} as for the computation of the base template $\mathcal{T}^a$, in this case 50\Mpch{}. The deep curve is obtained by considering a sample of galaxies twice deeper for galaxies $\mathbfit{x}_i$, thus 100\Mpch{}, than for deriving $\mathcal{T}^a$. }
\end{figure}

The above expression is exact up to the approximation of the linear perturbation theory. To go further we have to approximate the behaviour of $n_e(\mathbfit{r})$. In the following, we will make the following approximation: the galaxies have a much smaller size than the considered scales, their distances are large compared to their size, and their density profile is going to be simplified to the maximum. We consider that it can be expanded like:
\begin{equation}
  n_e(\mathbfit{r}) \simeq \sum_{i=1}^{N_\text{galaxy}} W(\mathbfit{r}-\mathbfit{x}_i),
\end{equation}
with $\mathbfit{x}_i$ the position of the $i$-th galaxy. This sum runs over all galaxy of the observable Universe. 
$W$ being a function with compact support, we approximate it as a small cone:
\begin{equation}
  W(\mathbfit{r}-\mathbfit{x}_i) \simeq A_e H(r-x_i) \Theta_i(\hat{r}),
\end{equation}
with $\Theta_i(\hat{r})=1$  if $\cos^{-1}(\hat{r}\cdot \hat{x}_i) \le R/x_i$ and zero otherwise, and $H(r)=1$ only if $|r| \le R$, zero otherwise, and $A_e$ the mean number density of electrons in the galaxy. This representation of the distribution of electrons in a galaxy is very schematic but sufficient to derive the window function. 
We now replace the expression $n_e(\mathbfit{r})$ in Equation~\eqref{eq:W_attempt0}:
\begin{align}
  \widetilde{\mathcal{W}}^{a,b}(k) & \simeq 4 \sum_{\ell,m} \sum_{\ell',m'} \frac{1}{C_\ell C_{\ell'}} g^a_{\ell,m} g^b_{\ell',m'}  \sum_{i,j=1}^{N_\text{galaxies}} \Bigg[ \int_{s=0}^{+\infty} \int_{s'=0}^{+\infty} \mathrm{d}^2 \hat{s} \mathrm{d}^2\hat{s'}\;  W(\mathbfit{s}-\mathbfit{x}_i) W(\mathbfit{s}'-\mathbfit{x}_j) Y_{\ell,m}(\hat{s}) Y_{\ell',m'}(\hat{s}') \hat{s}^c \hat{s'}^{d}  \\ 
 & \int\frac{3 \mathrm{d}^2\hat{k}}{4\pi} \hat{k}^c \hat{k}^d \mathrm{e}^{i \mathbfit{k}\cdot (\mathbfit{s}-\mathbfit{s'})} \Bigg], \\
  & = 16 R^2 \sum_{\ell,m} \sum_{\ell',m'} \frac{1}{C_\ell C_{\ell'}} g^a_{\ell,m} g^b_{\ell',m'}  \sum_{i,j} \int  \mathrm{d}^2 \hat{s} \mathrm{d}^2 \hat{s'} Y_{\ell,m}(\hat{s}) Y_{\ell',m'}(\hat{s}')  \hat{s}^c \hat{s'}^{d}  \int\frac{3 \mathrm{d}^2\hat{k}}{4\pi} \hat{k}^c \hat{k}^d h_i(\mathbfit{k}\cdot\hat{s}) h^{*}_j(\mathbfit{k}\cdot\hat{s}') \Theta_i(\hat{s}) \Theta_j(\hat{s}'),
\end{align}
with $i$ and $j$ running over the galaxies of the observable Universe, and
\begin{equation}
  h_i(k) = \sinc(k R) \mathrm{e}^{i k x_i}.
\end{equation}
We only consider scales much larger than the size of a galaxy, $R$, thus $k R \ll 1$ and $\sinc(k R)\simeq 1$ to first order in $k R$.
We find
\begin{align}
  \widetilde{\mathcal{W}}^{a,b}(k) & = 16 R^2 \sum_{\ell,m} \sum_{\ell',m'} \frac{1}{C_\ell C_{\ell'}} g^a_{\ell,m} g^b_{\ell',m'}  \sum_{i,j} \int  \mathrm{d}^2 \hat{s} \mathrm{d}^2 \hat{s'} Y_{\ell,m}(\hat{s}) Y_{\ell',m'}(\hat{s}')  \hat{s}^c \hat{s'}^{d}  \int\frac{3 \mathrm{d}^2\hat{k}}{4\pi} \hat{k}^c \hat{k}^d \Theta_i(\hat{s}) \Theta_j(\hat{s}') \mathrm{e}^{i \mathbfit{k}\cdot(\hat{s} x_i - \hat{s}' x_j)},\\
  & \simeq 16 \pi^2 R^2 \sum_{i,j} \sum_{\ell,m} \sum_{\ell',m'} \frac{1}{C_\ell C_{\ell'}} g^a_{\ell,m} g^b_{\ell',m'} Y_{\ell,m}(\hat{x}_i) Y_{\ell',m'}(\hat{x}_j) \hat{x}_i^c \hat{x}_j^d \left(\frac{R^2}{x_i x_j}\right)^2 \int\frac{3 \mathrm{d}^2\hat{k}}{4\pi} \hat{k}^c \hat{k}^d  \mathrm{e}^{i \mathbfit{k}\cdot(\mathbfit{x}_i - \mathbfit{x}_j)} \\
  & = 16\pi^2 R^2 \sum_{i,j} \mathcal{T}^a(\hat{x}_i) \mathcal{T}^b(\hat{x}_j)  \left(\frac{R^2}{x_i x_j}\right)^2 \hat{x}_i^c \hat{x}_j^d G^{c,d}\left(k(\mathbfit{x}_i-\mathbfit{x}_j)\right) \label{eq:window_prefinal}
\end{align}
with
\begin{align}
  G^{a,b}(\mathbfit{x}) & = G_0(|\mathbfit{x}|)\delta^{a,b} + G_1(|\mathbfit{x}|) \hat{x}^a \hat{x}^b \\
     G_0(x) &= 3 \frac{\sin x - x \cos x}{x^3} \\
     G_1(x) &= \frac{x^2 \sin x + 3 x \cos x - 3 \sin x}{x^3}
\end{align}
and
\begin{equation}
  \mathcal{T}^a(\hat{n}) = \sum_{\ell,m} \frac{1}{C_\ell} g^a_{\ell,m} Y_{\ell,m}(\hat{n}).
\end{equation}
The summation over all galaxies can effectively be truncated to the depth of the survey $R_S$. In Eq.~\eqref{eq:window_prefinal} the summation over the galaxies corresponds to compute the cross correlation between the weighted templates $\mathcal{T}^a$ and the projected density of the Universe of the sky given by the other terms. This correlation small, though not negligible, whenever the galaxy $i$ or $j$ is outside the volume on which we are fitting. We show the amplitude of this effect in Figure~\ref{fig:test_convergence}. We note that the function is reduced at $k \ga 0.1 h$~Mpc$^{-1}$. The shape is preserved for $k \la 0.1 h$~Mpc$^{-1}$.

We note that the derived window function cannot be trusted on scales $k \ga 1/R$. The reason lies in the simplistic assumption on the density profiles and the approximation on the expansion of the function $\sinc$.

The last missing piece consists in deriving the covariance matrix $\mathcal{M}^{a,b}$. As we have used an approximation for only one part of the derivation of the unweighted window function $\widetilde{\mathcal{W}}^{a,b}$, we ought to do the same for the derivation of $\mathcal{M}^{a,b}$ even though it is not required. The covariance matrix becomes:
\begin{align}
  \mathcal{M}^{a,b} & = 2 \sum_{\ell,m} \frac{1}{C_\ell} g^a_{\ell,m} \left(\int\text{d}^2\hat{s} \int_{s=0}^{R_S} \text{d}s Y_{\ell,m}(\hat{s}) n_e(s\hat{s}) \hat{s}^b\right) \\
   & \simeq 4 \pi R \sum_{i, x_i \le R_S} \sum_{\ell,m} \frac{1}{C_\ell} g^a_{\ell,m} \left(\frac{R}{x_i}\right)^2 Y_{\ell,m}(\hat{x}_i) \hat{x}_i^b \\
  & = 4\pi R \sum_{i / x_i \le R_S} \mathcal{T}^a(\hat{x}_i) \left(\frac{R}{x_i}\right)^2 \hat{x}_i^b, \label{eq:M_final}
\end{align}
with the summation is limited to galaxies with $x_i \le R_S$ as indicated by $i / x_i \le R_S$. We note that we lose the obvious symmetry of $\mathcal{M}^{a,b}$ by making this approximation. Using the expression \eqref{eq:M_final}, we note that
\begin{equation}
  \widetilde{\mathcal{W}}^{a,b}(k=0) = \sum_{c=1}^3 \mathcal{M}^{a,c} \mathcal{M}^{c,b}.
\end{equation}
Consequently the weighted window function $\mathcal{W}^{a,b}(k)$, defined through the covariance
\begin{equation}
  \langle V^a V^b \rangle = a^2 H^2 f^2 \int_{k=0} \frac{\mathrm{d} k}{6\pi^2} P(k) \mathcal{W}^{a,b}(k) =  \sum_{c,d=1}^3 \left(\mathcal{M}^{-1}\right)^{a,c} \left(\mathcal{M}^{-1}\right)^{d,b} \langle \mathcal{V}^c \mathcal{V}^d \rangle
\end{equation}
is related to $\widetilde{\mathcal{W}}^{a,b}$ through:
\begin{equation}
  \mathcal{W}^{a,b}(k) = \sum_{c,d=1}^3 \left(\mathcal{M}^{-1}\right)^{a,c} \left(\mathcal{M}^{-1}\right)^{d,b} \widetilde{\mathcal{W}}^{c,d}(k).
\end{equation}
We conclude by acknowledging that $\mathcal{W}^{a,b}(k=0) = \delta^{a,b}$.

\section{Numerical test on mock sky maps}
\label{app:numerical_test}

\begin{figure}
	\begin{center}
		\includegraphics[width=0.7\hsize]{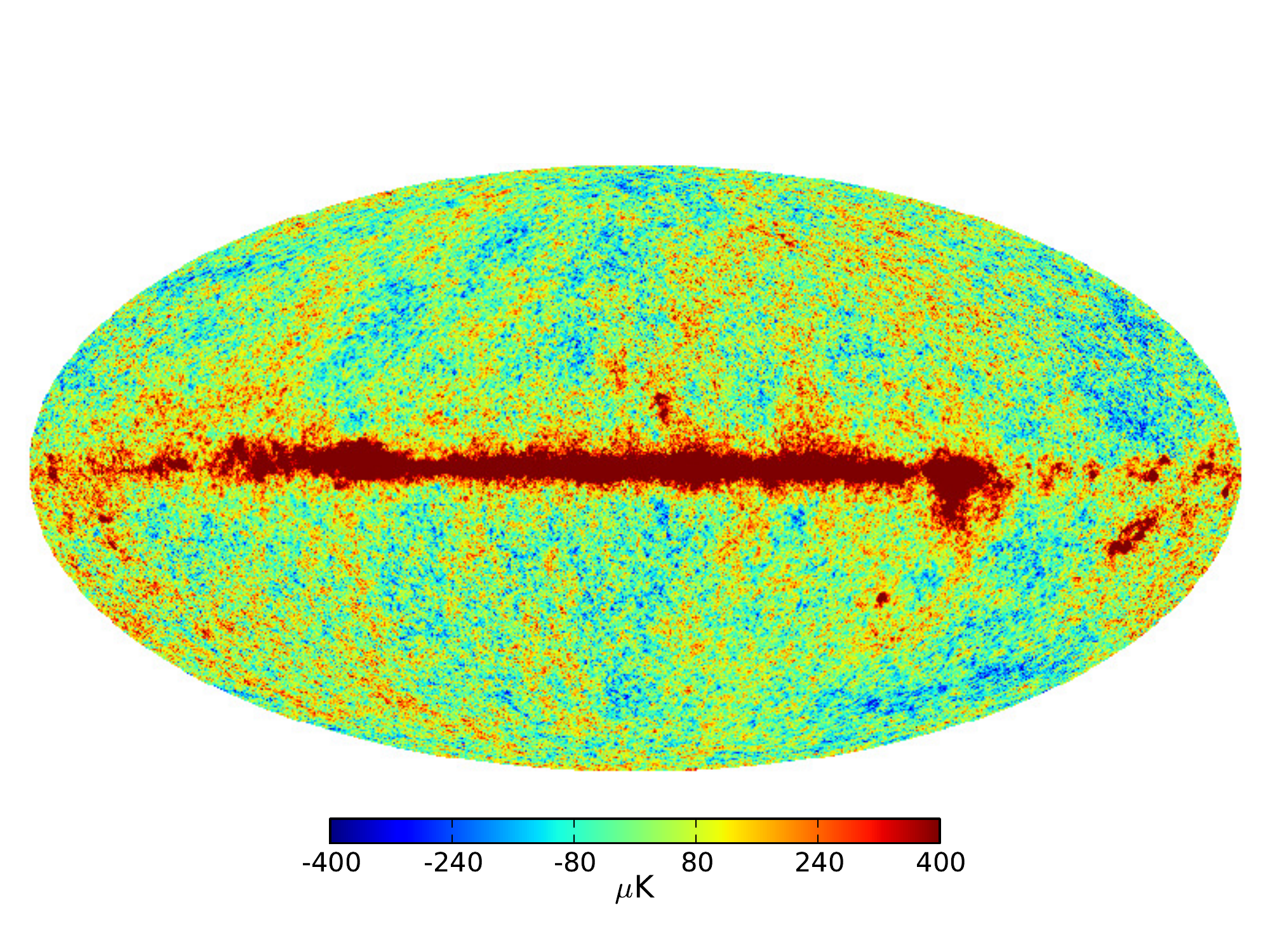}
	\end{center}
\caption{\label{fig:mock_sky} Mock CMB sky map generated as indicated in Appendix~\ref{app:numerical_test}. We show  the mock sky map obtained in the $W$ band, at $N_\text{side}=512$, assuming the characterization of the statistical properties of the map given by the WMAP collaboration and our best fit value for the templates that we have adopted. }
\end{figure}

We have checked using mock sky maps our numerical implementation of our template matching likelihood of Section~\ref{sec:stat_method}. We have generated mock sky maps using the best fit value for our model which are given in the Tables~\ref{tab:foregrounds} (line one) and \ref{tab:bulk_flows} (line one). We have used the best fit value obtained using WMAP7 data for the parameters of the $\Lambda$CDM model. We have generated randomly the phases of the primary CMB fluctuations. All the components (tSZ, kSZ, point sources, galaxy foreground contamination, CMB) have then been summed, smoothed using the beam transfer function provided by the WMAP collaboration for the 7-year data. In the end, we have added a random noise, uncorrelated in pixel space, and weighed according to the number of observations. We produce in such a way one mock sky map per frequency channel (Q, V and W). The primary CMB is kept exactly the same between sky maps but the noise is independent. We show in Figure~\ref{fig:mock_sky} the final mock sky map that we have obtained for the W channel.

\begin{figure}
  \begin{center}
    \includegraphics[width=.5\hsize]{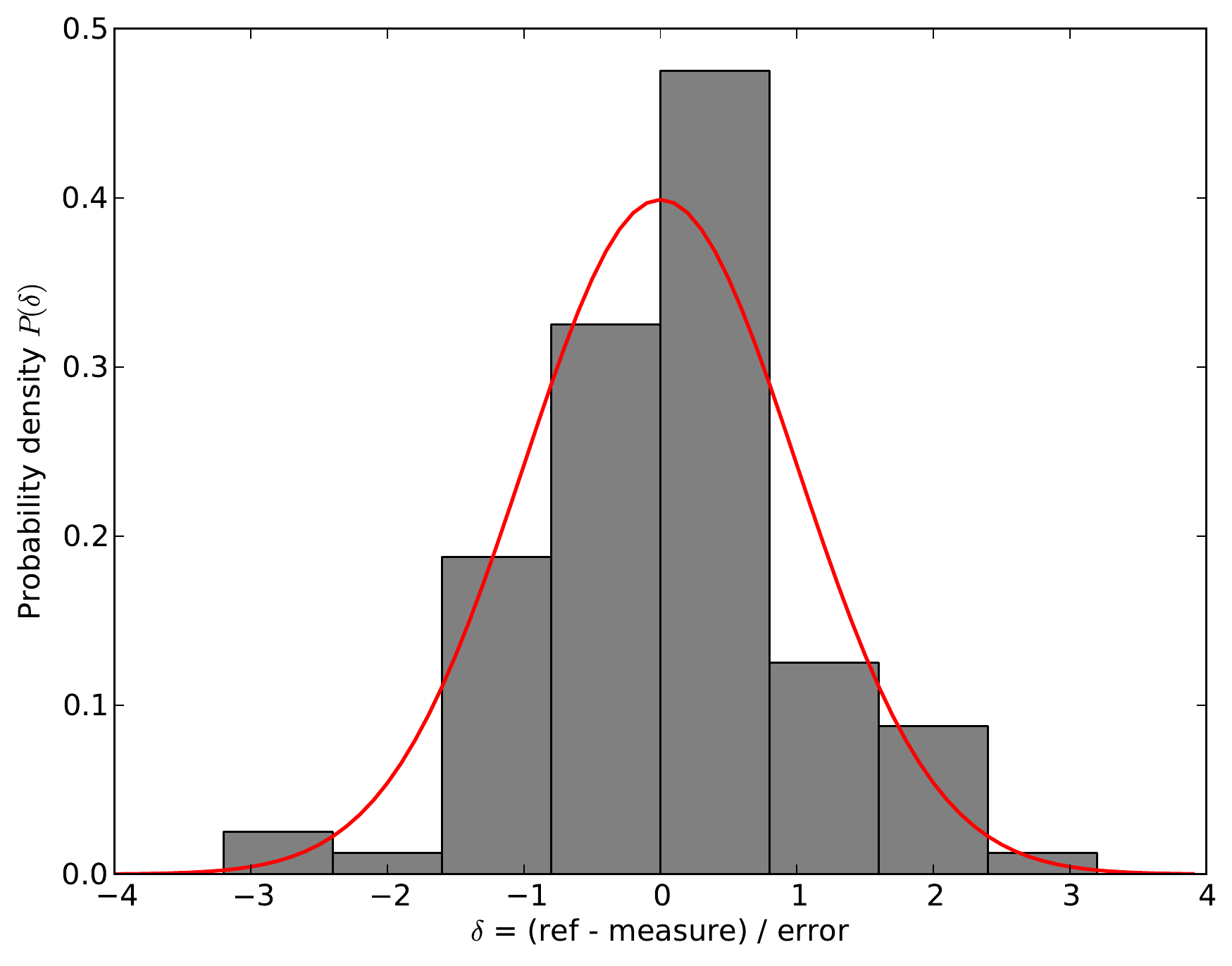}
  \end{center}
  \caption{\label{fig:posterior_ksz} Example of the full posterior of the X component of the kSZ bulk flow signal detected in 100 simulations mimicking WMAP maps characteristics, as explained in Section~\ref{app:numerical_test}. The histograms are derived from binning the results of the mock measurements on simulations, after normalization by the predicted standard deviation. The red solid curve is the expected Gaussian curve, with unit variance.}
\end{figure}

Table~\ref{tab:mocksky_test} gives the input value that we have used to generate the mock sky maps, then the best-fitted value with their attach predicted error bars by the Bayesian analysis for one mock sky map. Most of the fitted values are in agreement with the input value within the $2\sigma$ range. The exception being the fitted values of the foreground $T_K-T_{K_A}$ which are at $2.5\sigma$ from the input value. Finally in Figure~\ref{fig:posterior_ksz}, we show the distribution for the mean of the X-component of the kSZ bulk flow for 100 mock-sky. The distribution is centered on the actual input value that we have chosen and normalised by the standard deviation predicted by our algorithm. We have overplotted, with a thick solid red line, the expected distribution for these values, which is a Gaussian distribution centered on zero with unit variance. During the tests, we have also noted a small numerical bias in the computation of the amplitude of the tSZ template. At the level of present error bars, this is not significant ($\sim 0.2-0.5\sigma$) and it can be significantly reduced by increasing the resolution of the maps. We thus conclude that our implementation of the statistical method given in Section~\ref{sec:stat_method} works as advertised, and provided the models are correct, is unbiased.

\begin{table}
	\begin{center}
	\begin{tabular}{cccc}
		\hline
		Parameter & Frequency channel & Input value & Posterior value \\
		\hline
		\hline
		Monopole & Q & 0.40~\mK & $0.46 \pm 0.03$~\mK \\
		 & V & 0.19~\mK & $0.25 \pm 0.03$~\mK \\
		 & W & 0.11~\mK & $0.17 \pm 0.03$~\mK\\
		\hline
		Dipole (1,0) & Q& 10~\uK& $12.4\pm 1.6$~\uK\\
		 & V & 9~\uK & $11.5\pm 1.6$~\uK\\
		 & W & 1~\uK & $3.60\pm 1.60$~\uK\\
		\hline
		Dipole (1,1) & Q& $(-0.3-6i)$~\uK & $(6\pm 7)-(3\pm 5)i$~\uK \\
		 & V & $(0.2-4i)$~\uK & $(6\pm 7)-(0.4 \pm 5)i$~\uK \\
		 & W & $(0.4-6i)$~\uK & $(6\pm 7)-(2 \pm 5)i$~\uK \\
		\hline
		$H_\alpha$ & Q & $1.5$~\muKR & $2.2\pm 0.4$~\muKR \\
		 & V & $1.0$~\muKR & $1.7 \pm 0.4$~\muKR \\
		 & W & $0.7$~\muKR & $1.4 \pm 0.4$~\muKR \\
		\hline
		Dust & Q & 0.1 & $-0.3\pm 0.3$\\
		 & V & 0.3 & $-0.1 \pm 0.3$\\
		 & W & 1.0 & $0.6 \pm 0.3$ \\
		\hline
		$T_K-T_{K_A}$ & Q & 0.37 & $0.42\pm 0.02$ \\
		 & V & 0.18 & $0.23 \pm 0.02$ \\
		 & W & 0.11 & $0.16 \pm 0.02$ \\
		\hline
		PSC & Q& 60~\mJy & $60\pm 20$~\mJy \\
		 & V & 120~\mJy & $80\pm 60$~\mJy\\
		 & W & 400~\mJy & $500 \pm 300$~\mJy \\
		\hline
		tSZ & & 1.0 & $0.93\pm 0.12$ \\
		kSZ $V_x$ & & 559~\kms & $462 \pm 267$~\kms \\
		kSZ $V_y$ & & -412~\kms & $-431\pm 286$~\kms \\
		kSZ $V_z$ & & -93~\kms & $-76 \pm 187$~\kms \\
		\hline
	\end{tabular}
	\end{center}
	\caption{\label{tab:mocksky_test} Result of the test on a single mock-sky realization. }
  \parbox{\hsize}{
  {\sc Note}: We have used some fiducial realistic values for the model mentioned in the table. These values are derived from best-fitting the models on WMAP data. The posterior values is derived from the mean and standard deviation as provided by the posterior distribution.
  }
\end{table}

\end{document}